\documentclass{emulateapj}

\newcommand\gsim{\,\lower3pt\hbox{$\sim$}\llap{\raise2pt\hbox{$>$}}\,}
\newcommand\lsim{\,\lower3pt\hbox{$\sim$}\llap{\raise2pt\hbox{$<$}}\,}

\usepackage{graphicx}
\usepackage{epsfig}
\usepackage{subfigure}
\usepackage{natbib}

\shortauthors{LUGAZ ET AL.}

\shorttitle{A COMPLEX EVENT OBSERVED BY SECCHI}

\begin{document}

%
%

\title{The Deflection of the Two Interacting Coronal Mass Ejections of 2010 May 23-24 as Revealed by Combined In situ Measurements and Heliospheric Imaging}

\author{N.\ Lugaz\altaffilmark{1}, C.~J.~Farrugia\altaffilmark{1}, J.~A. Davies\altaffilmark{2}, C. M{\"o}stl\altaffilmark{3,4,5}, C.~J.~Davis\altaffilmark{2,6},  I.~I.\ Roussev\altaffilmark{7,8}, M.~Temmer\altaffilmark{4}}
\altaffiltext{1}{Space Science Center, University of New Hampshire, Durham, NH, USA}
\altaffiltext{2}{RAL Space, STFC Rutherford Appleton Laboratory, Chilton, UK}
\altaffiltext{3}{Space Science Laboratory, University of California, Berkeley, CA, USA}
\altaffiltext{4}{Kanzelh\"ohe Observatory-IGAM, Institute of Physics, University of Graz, Graz 8010, Austria}
\altaffiltext{5}{Space Research Institute, Austrian Academy of Science, Graz 8042, Austria}
\altaffiltext{6}{Department of Meteorology, University of Reading, Berkshire, UK}
\altaffiltext{7}{Yunnan Astronomical Observatory, Chinese Academy of Sciences, Kunming 650011, Yunnan, China}
\altaffiltext{8}{Institute for Astronomy, University of Hawaii, Honolulu, HI, USA}

%
%

\begin{abstract}
In 2010 May 23-24, SDO observed the launch of two successive coronal mass ejections (CMEs), which
were subsequently tracked by the SECCHI suite onboard STEREO. 
Using the COR2 coronagraphs and the heliospheric imagers (HIs), the initial direction of both CMEs is determined to be slightly west of the Sun-Earth line. We derive the CME kinematics, including the evolution of the CME expansion until 0.4 AU. We find that, during the interaction, the second CME decelerates from a speed above 500 km~s$^{-1}$ to 380~km~s$^{-1}$, the speed of the leading edge of the first CME. STEREO observes a complex structure composed of two different bright tracks in HI2-A but only one bright track in HI2-B. 
{\it In situ} measurements from {\it Wind} show an ``isolated'' ICME, with the geometry of a flux rope preceded by a shock. Measurements in the sheath are consistent with draping around the transient.   
By combining remote-sensing and {\it in situ} measurements, we determine that this event shows a clear instance of deflection of two CMEs after their collision, and we estimate the deflection of the first CME to be about 10$^\circ$ towards the Sun-Earth line. The arrival time, arrival speed and radius at Earth of the first CME are best predicted from remote-sensing observations taken before the collision of the CMEs. Due to the over-expansion of the CME after the collision, there are few, if any, signs of interaction in {\it in situ} measurements. This study illustrates that complex interactions during the Sun-to-Earth propagation may not be revealed by {\it in situ} measurements alone.

\end{abstract}

\section{INTRODUCTION} \label{intro}

The interaction of successive interplanetary coronal mass ejections (ICMEs) between the Sun and the Earth was first inferred from the analyses of multi-spacecraft {\it in situ} measurements during the 1980s, when the Helios satellites in the inner heliosphere complemented the ISEE spacecraft near 1~AU \citep[]{Burlaga:1987}. With the improvement of coronagraphic observations, it was confirmed in the 1990s that successive ICMEs can merge, as well as interact with solar wind streams to form a compound stream or complex ejecta \citep[]{Gopalswamy:2001, Burlaga:2002,Burlaga:2003, Wang:2002}. Complex ejecta are often associated at Earth with extended periods of strong southward $B_z$ \citep[e.g.][]{Wang:2003a,Farrugia:2006} and with intense geomagnetic storms \citep[]{Burlaga:1987, Farrugia:2006, Farrugia:2006b, Xie:2006}. Until the launch of the {\it Solar-Terrestrial Relations Observatory} \citep[STEREO, see:][]{Kaiser:2008} in 2006, there was very limited data available beyond the field-of-view (FOV) of coronagraphs ($\sim 0.15$~AU). In addition, {\it in situ} measurements outside of Earth's direct vicinity have been limited to planetary missions, which, generally, cannot easily be combined with measurements at 1~AU to study the interaction of successive CMEs and the formation of complex ejecta. Most of what we have learned about CME-CME interaction comes from numerical simulations. Such simulations were pioneered by \citet{Vandas:1997} and \citet{Schmidt:2004} and they have been performed since using 2.5-D and 3-D magneto-hydrodynamical codes \citep[]{Wu:2002, Odstrcil:2003, Lugaz:2005b,  Xiong:2006,Xiong:2009}. A few real events have also been simulated \citep[]{WuCC:2007, Lugaz:2007}.

\begin{figure*}[ht*]
\begin{center}
{\includegraphics*[width=5.4cm]{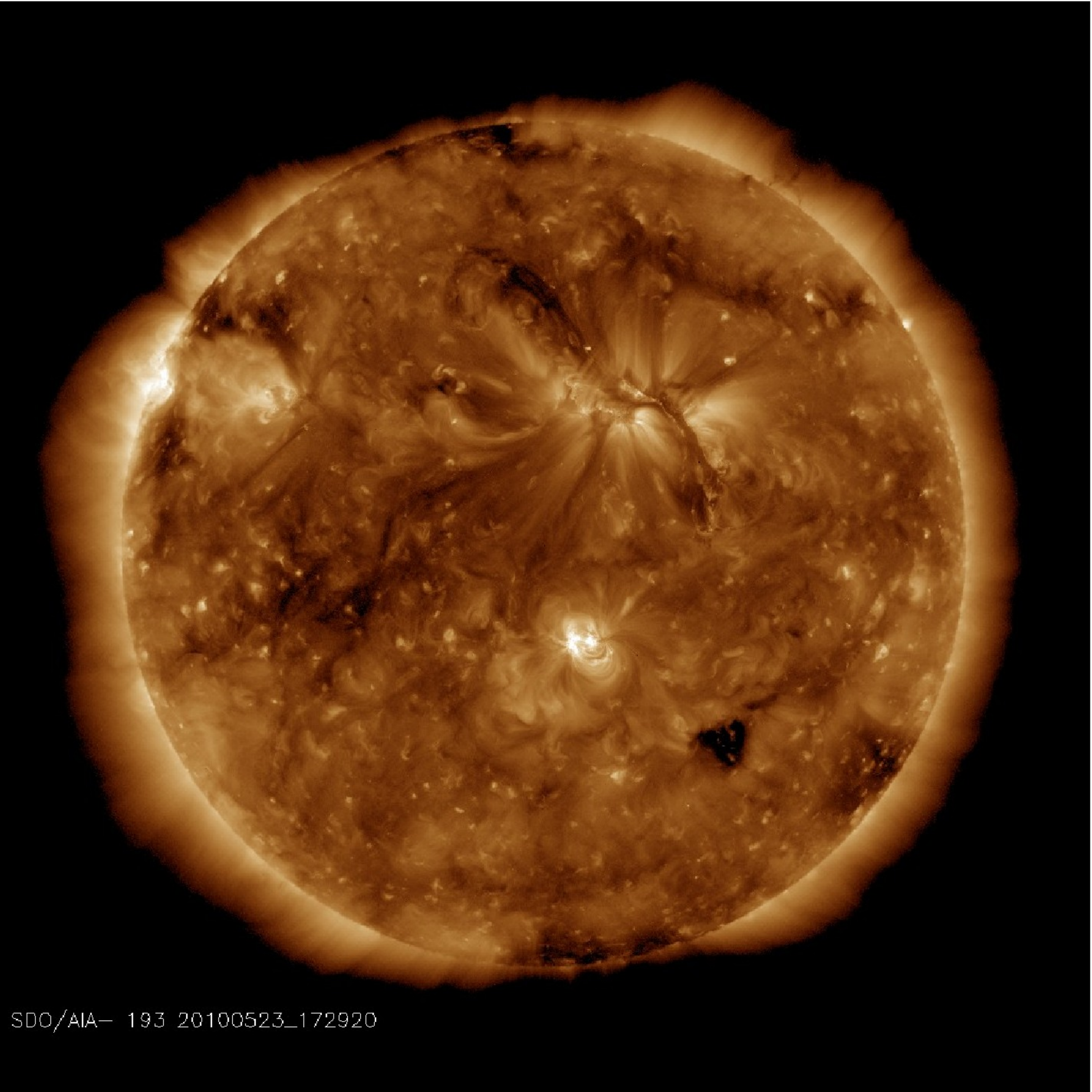}}
{\includegraphics*[width=5.4cm]{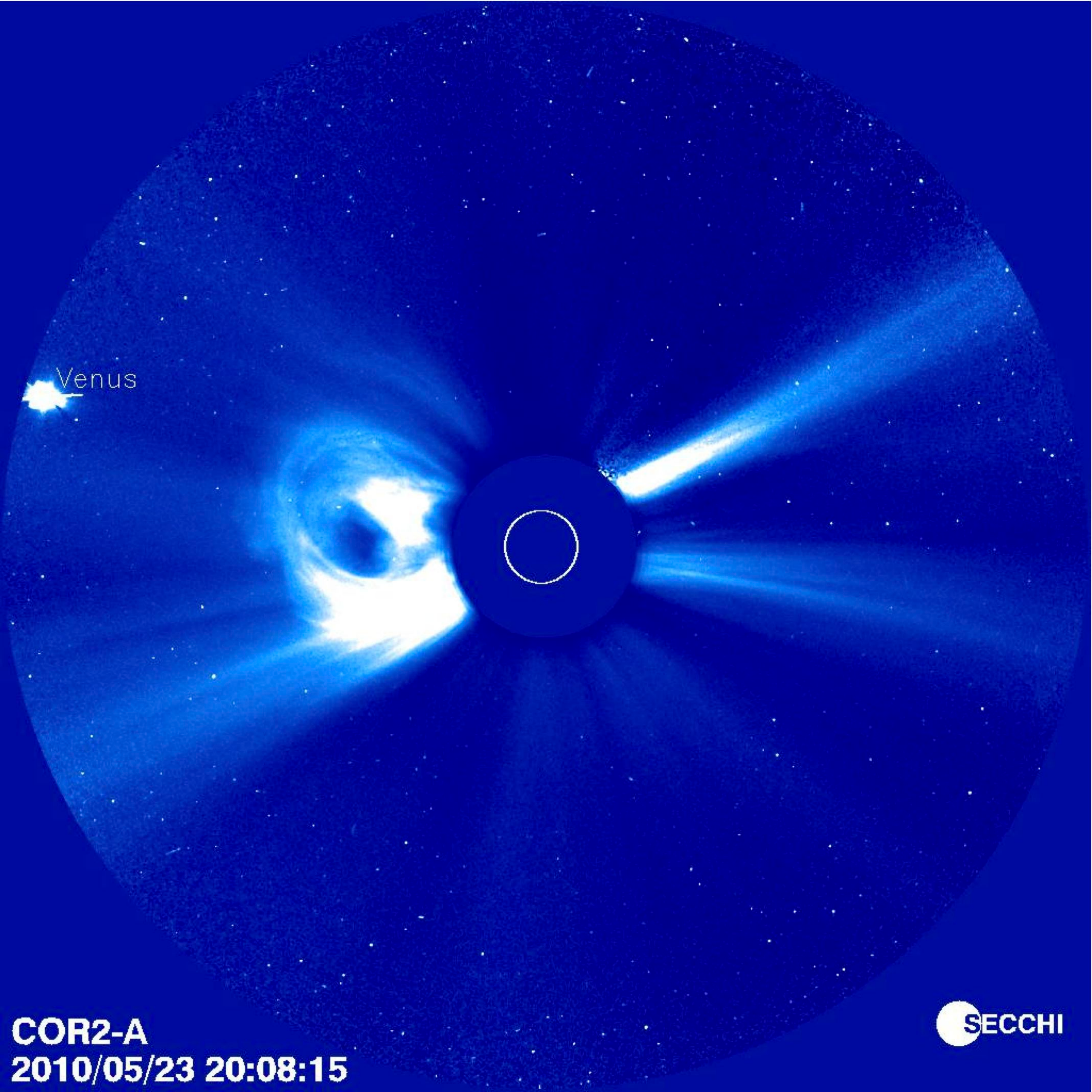}}
{\includegraphics*[width=5.4cm]{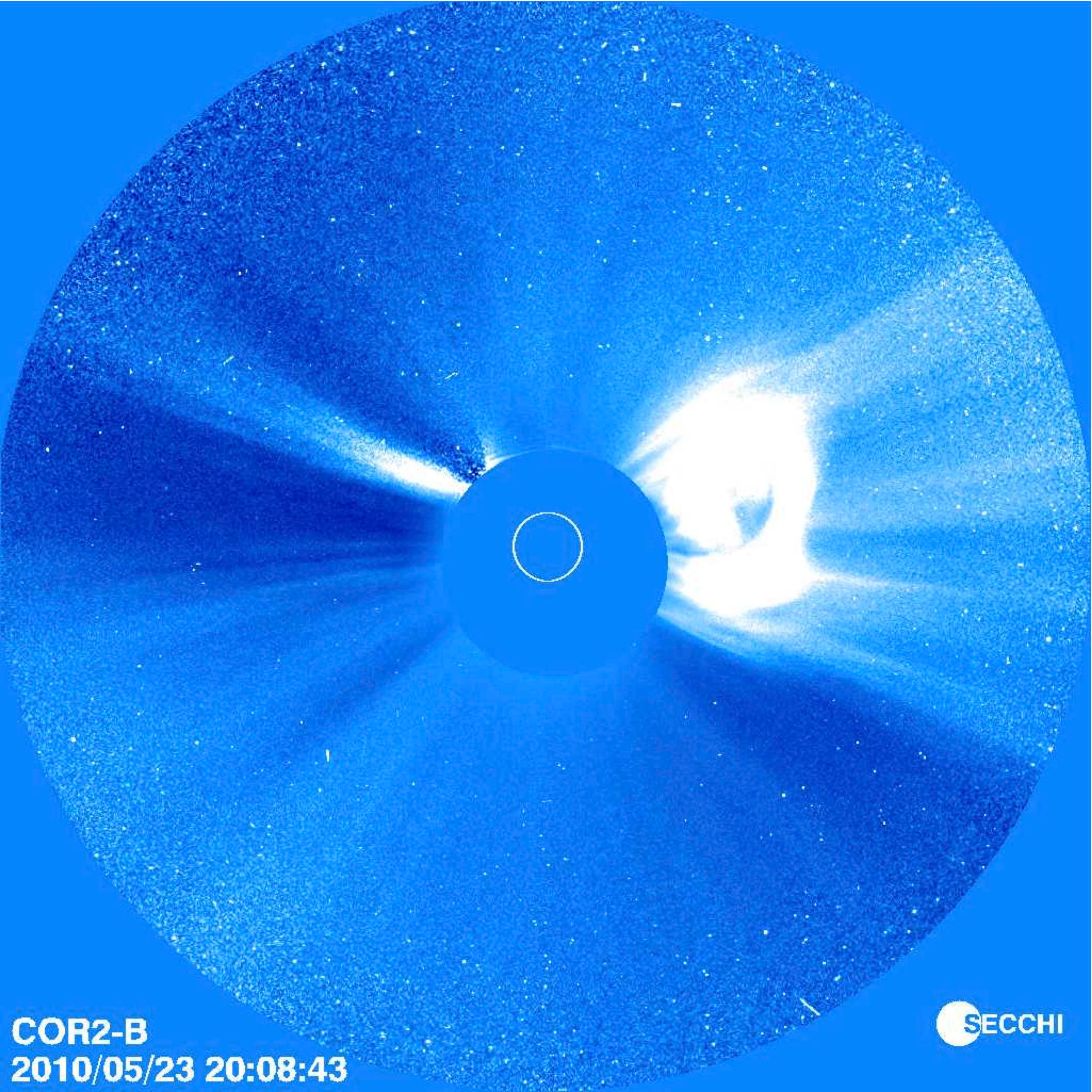}}\\
{\includegraphics*[width=5.4cm]{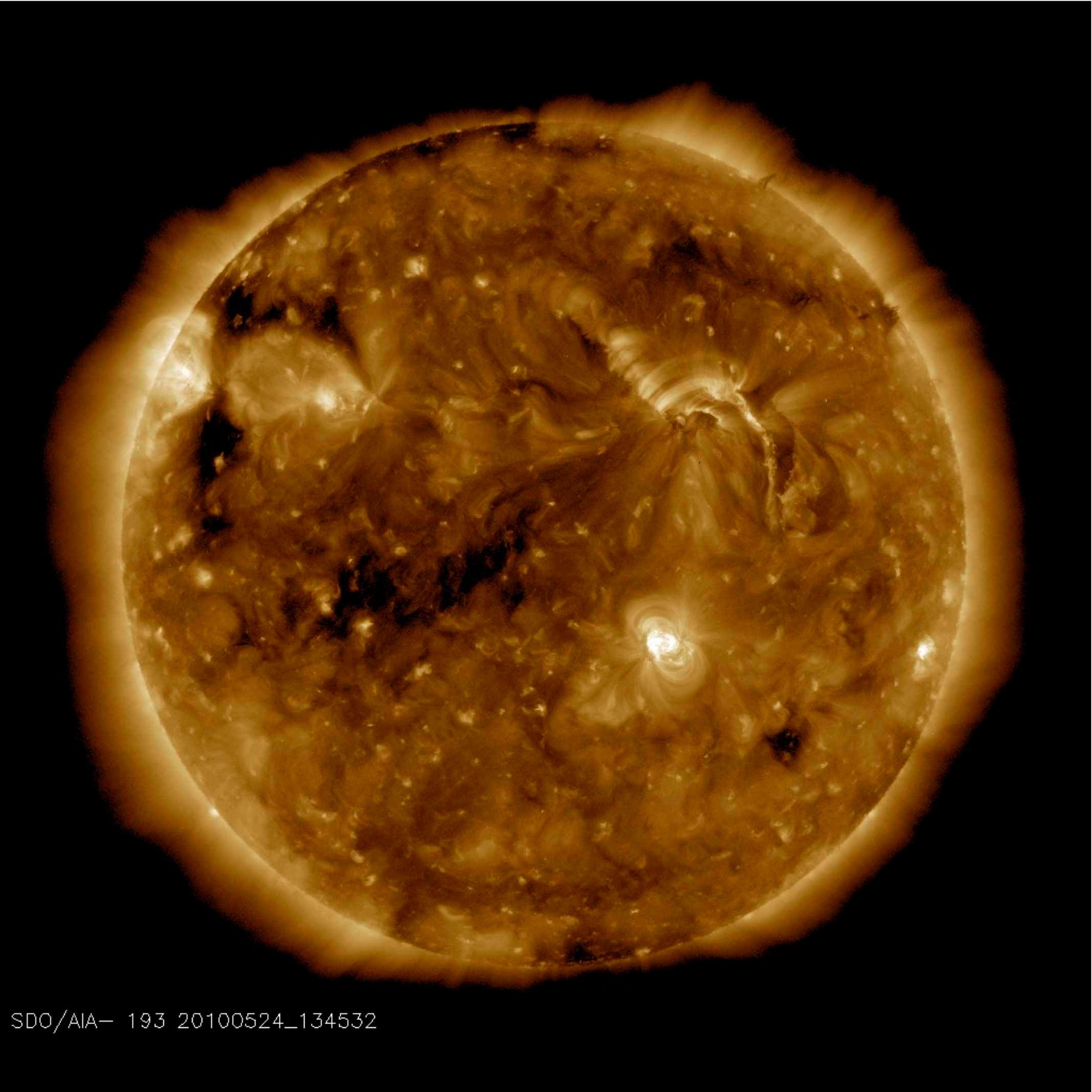}}
{\includegraphics*[width=5.4cm]{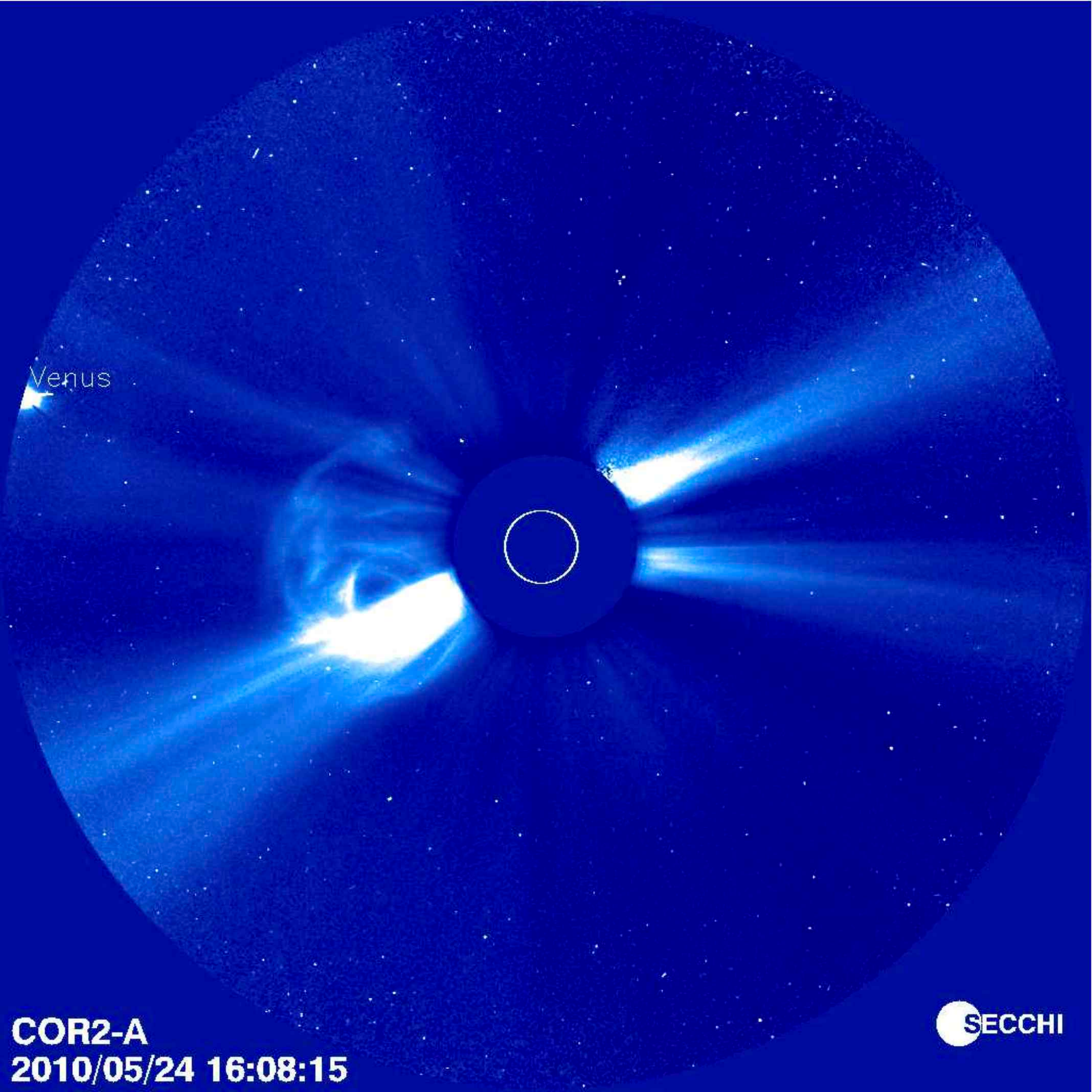}}
{\includegraphics*[width=5.4cm]{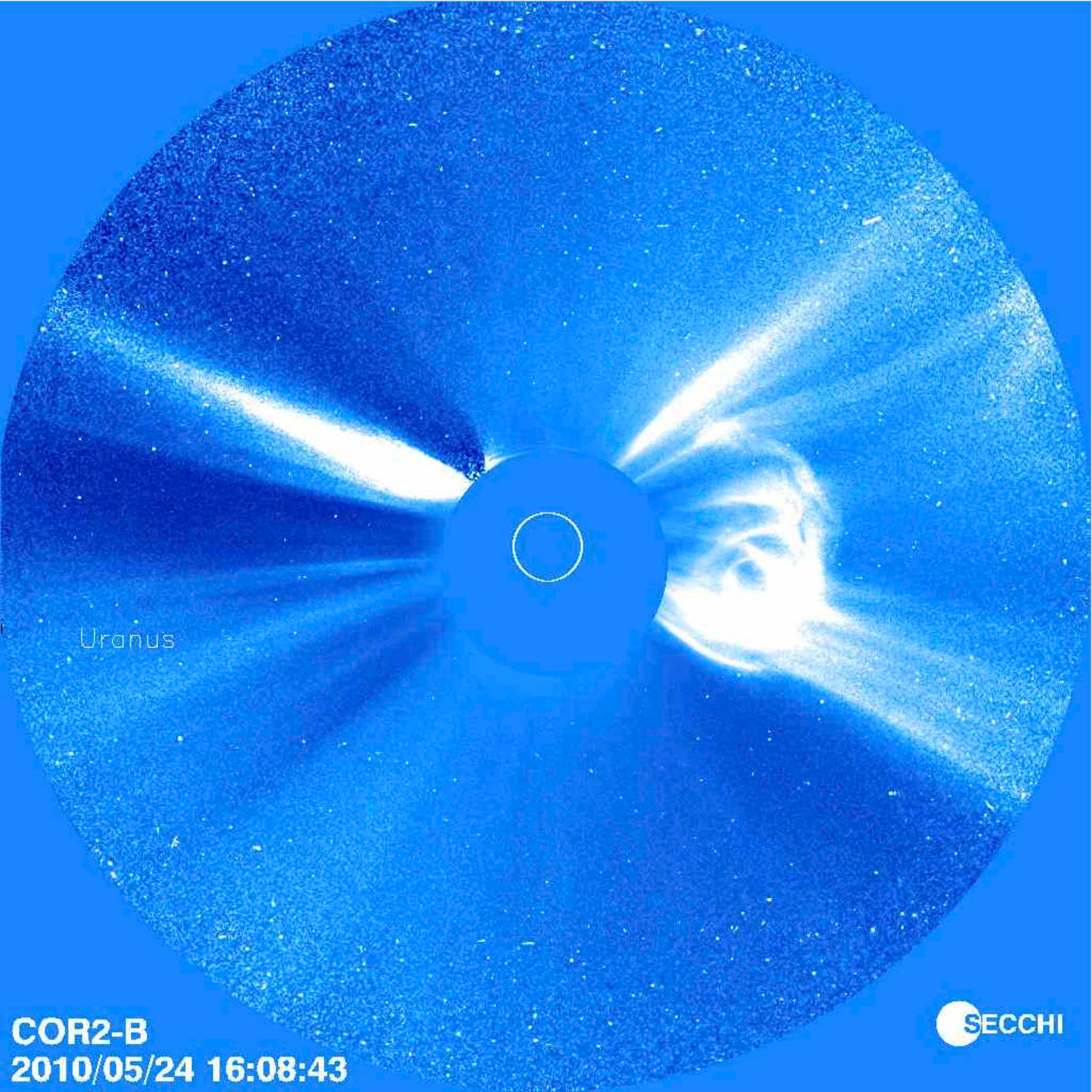}}
\caption{CMEs of 2010 May 23 (CME1, top) and 2010 May 24 (CME2, bottom) as observed by SDO 193~\AA~(left), COR-2 onboard STEREO-A (middle) and STEREO-B (right).}
\end{center}
\end{figure*}

Based on these simulations, it was found that the interaction of successive CMEs can cause intervals of southward $B_z$ with longer duration and larger magnitude \citep[]{Wang:2003a, Lugaz:2008a,  Xiong:2009} and also increase the density in the sheath \citep[]{Lugaz:2005b}, which could have important consequences for geo-effectiveness \citep[]{Farrugia:2006}. Some of the main remaining areas of investigation related to CME-CME interaction are: (i) the momentum exchange between the successive eruptions during their interaction, (ii) the fate of the related shocks, (iii) the possibility of CME-CME ``cannibalism'' through total reconnection of one of the CMEs, and, (iv) the deflection and rotation of a CME during the interaction. Regarding the momentum exchange during the interaction, it has been proposed that the interaction is elastic, perfectly inelastic, slower than inelastic due to the presence of the MHD obstacle \citep[]{Temmer:2012}, super-elastic \citep[]{Wang:2012}, or that the momentum is transferred through the propagation of the shock inside the first CME (see discussions in \citet{Lugaz:2009c}, 
\citet{Farrugia:2006} and \citet{Xiong:2006}).

With the Sun-Earth Connection Coronal and Heliosperic Investigation (SECCHI) suite \citep[]{Howard:2008} onboard STEREO, it is now possible to track the density fronts associated with CMEs from the corona to 1~AU \citep[e.g., see][]{Davies:2009}, to forecast the arrival at Earth of the ICME \citep[]{Davis:2011}. A particularly promising line of research is to combine these remote-sensing observations with {\it in situ} measurements from ACE, {\it Wind}, and the STEREO spacecraft to analyze the evolution of CMEs \citep[]{Wood:2009a, Moestl:2010, Liu:2010b, Rouillard:2010, Rollett:2012}. Because of the deep solar minimum encompassing most of the STEREO mission, there have not been many instances of successive and potentially interacting CMEs observed by SECCHI until early 2010. One major exception is the 2007 January 24-25 CMEs. This event has been fully analyzed in a number of articles, using STEREO and SMEI data \citep[]{Harrison:2009, Webb:2009, Lugaz:2009c} as well as numerical simulations \citep[]{Lugaz:2008b, Lugaz:2009b, Odstrcil:2009}. While this was an instance of interacting CMEs, the lack of {\it in situ} data and a long data gap make this event not particularly enlightening to understand CME-CME interaction. Another example of recent observations of  multiple CMEs is the series of eruptions in late July and early August 2010. These CMEs have been extensively studied \citep[]{Harrison:2012,Liu:2012,Temmer:2012}, but the number of involved CMEs (up to six, four of which occurred in one day) and the complexity of the observations make these events not optimal for studying the physical processes underlying CME-CME interaction. 

\begin{figure*}[ht*]
\begin{minipage}[b]{0.5\linewidth}\centering
{\includegraphics*[width=6.5cm]{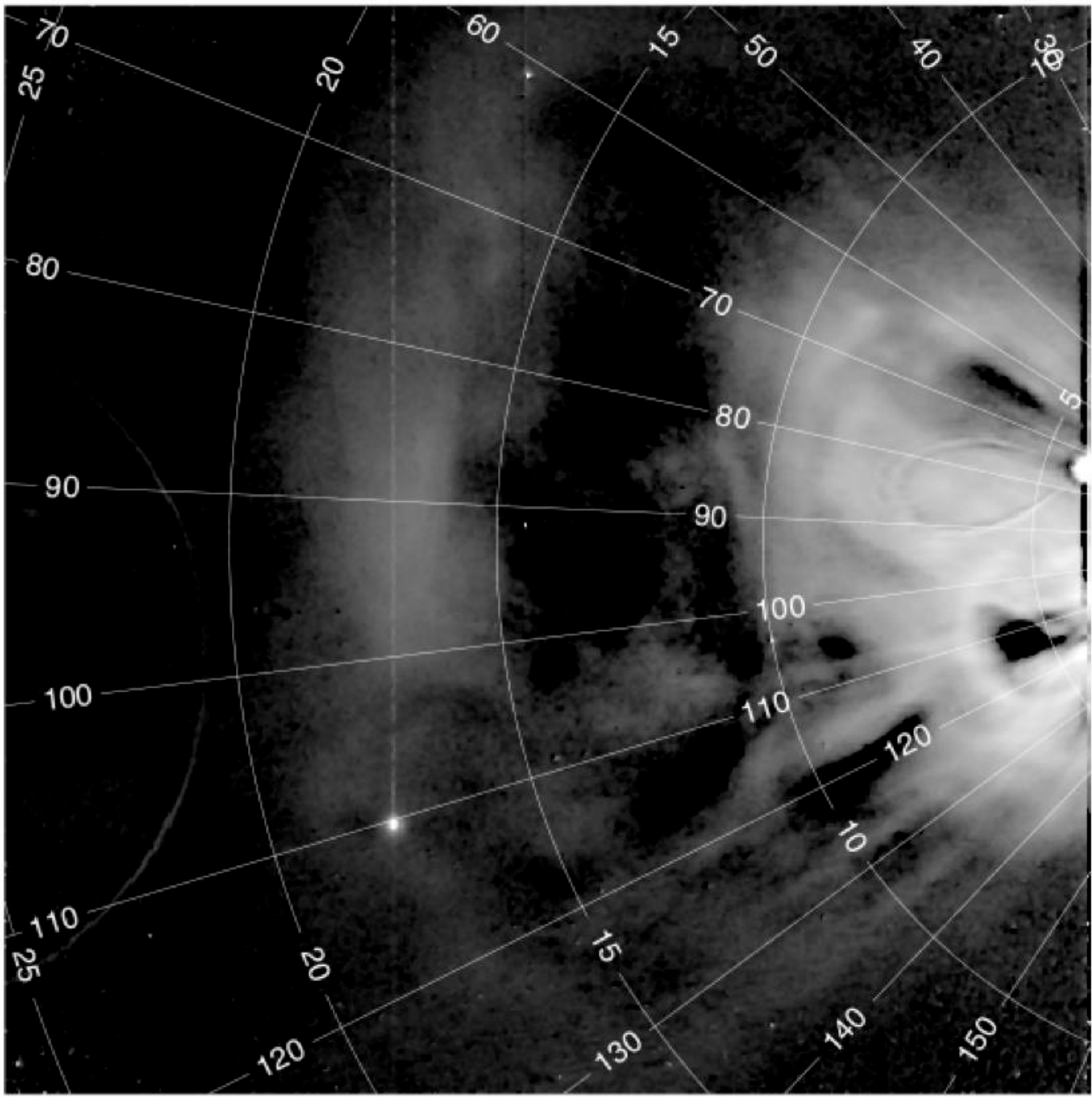}}
{\includegraphics*[width=6.5cm]{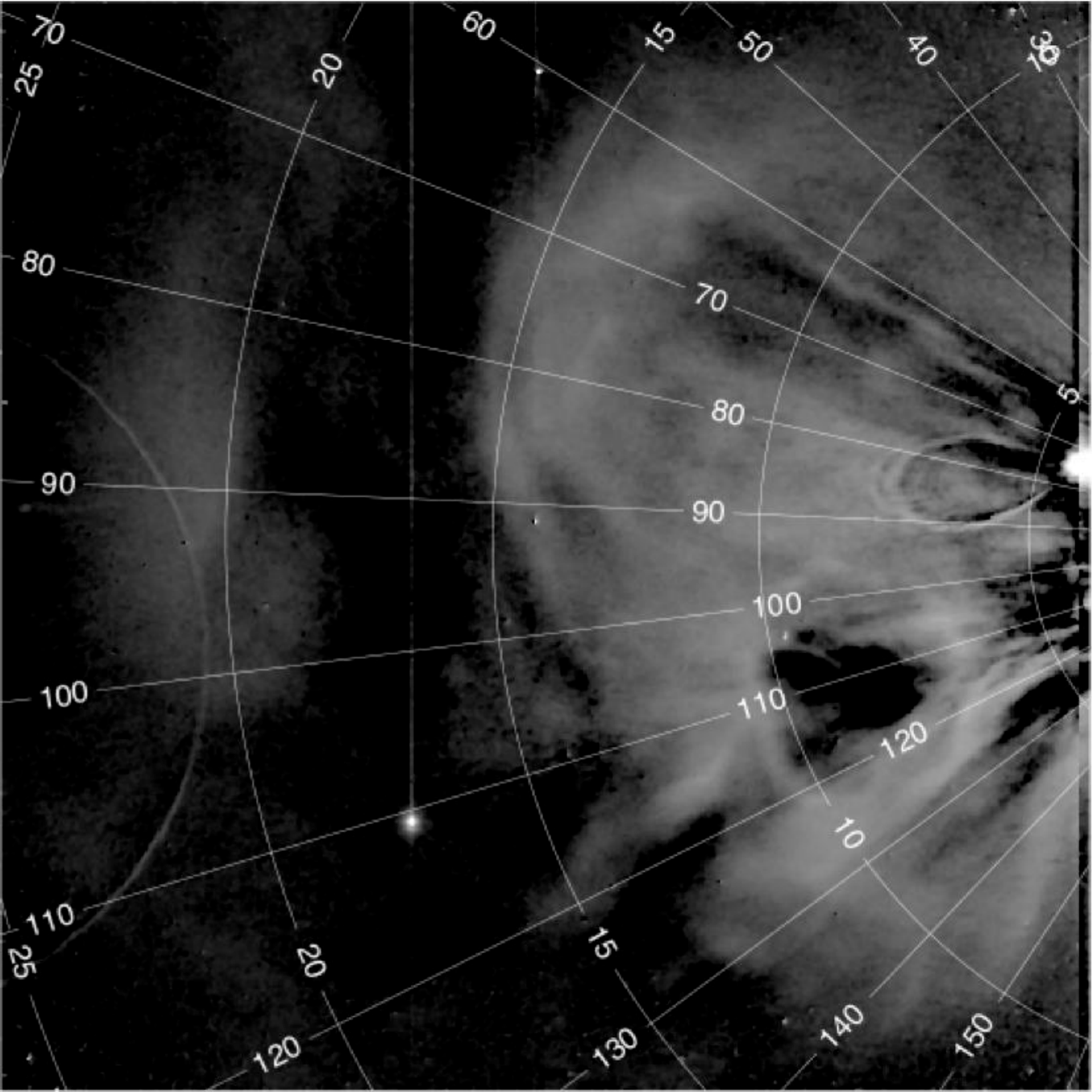}}
\end{minipage}
\begin{minipage}[b]{0.5\linewidth}\centering
{\includegraphics*[width=6.5cm]{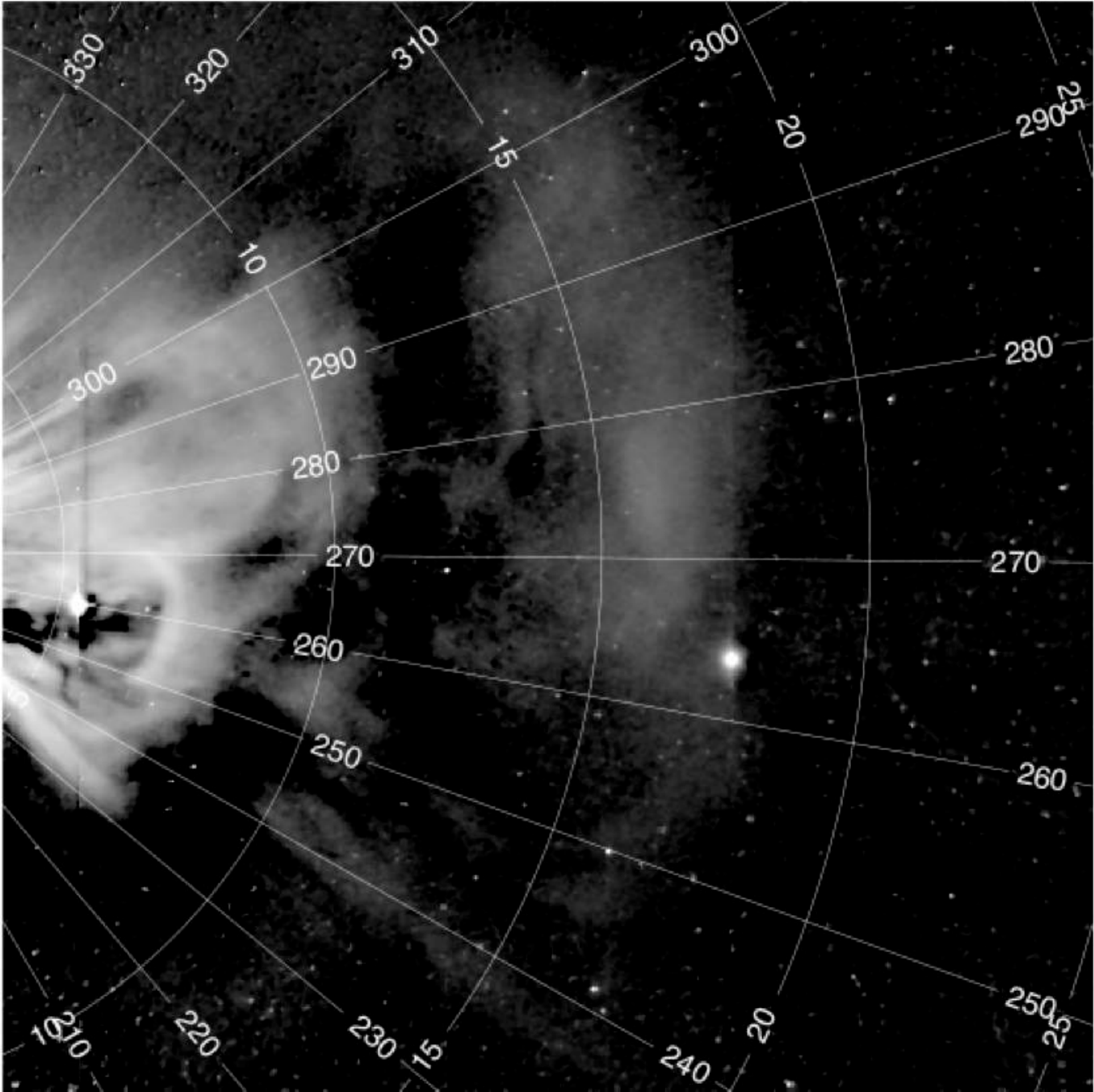}}
{\includegraphics*[width=6.5cm]{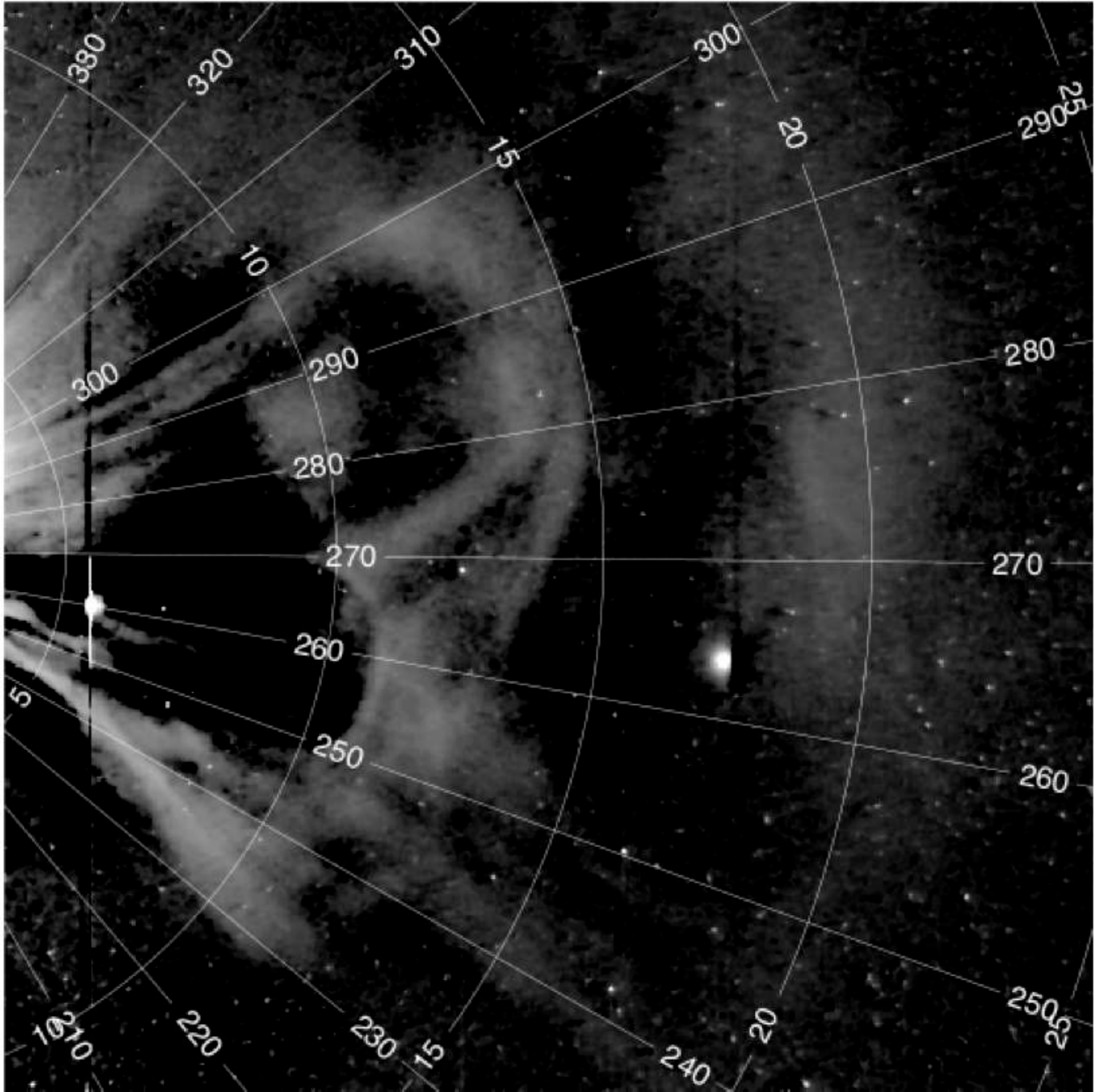}}
\end{minipage}
\caption{Background-subtracted images of the 2 CMEs on 2010 May 25 in HI1 onboard STEREO-A ({\it left}) and STEREO-B ({\it right}). The two rows correspond to 00:09 UT ({\it top})  and 06:49 UT ({\it bottom}). The concentric circles show the elongation angles and the radial lines the position angles, both in degrees. The position in elongation angle of CME1 is around 17$^\circ$ (top) and 21$^\circ$ (bottom); that of CME2 is around 10$^\circ$ (top) and 15$^\circ$ (bottom). [{\it This figure is available as two mpeg files in the electronic edition of the Journal}].}
\end{figure*}

In this paper, we present an analysis of observations by STEREO/SECCHI combined with measurements from the {\it Wind} spacecraft of two CMEs, which erupted on 2010 March 23 and 24. In section \ref{observation}, we discuss the remote-sensing observations of the two CMEs in extreme-ultraviolet (EUV) imagers, coronagraphs and heliospheric imagers \citep[HIs, see:][]{Eyles:2009}. In section \ref{direction}, we derive the CME directions using a number of methods. We analyze the measurements and derive the CME kinematics in the HIs FOVs in section \ref{HI1}, and we analyze the associated {\it in situ} measurements in section \ref{insitu}. We discuss our results and propose a coherent scenario for the CME-CME interaction in section \ref{discuss} and we conclude in section \ref{conclusion}.

\section{REMOTE-SENSING OBSERVATIONS} \label{observation}
\subsection{First CME}

On 2010 May 23, SDO/AIA and SOHO/EIT observed a B1 flare and a filament eruption originating from N19W12. STEREO-A was at a heliocentric distance of 0.956~AU and an angular separation of $71.5^\circ$ to the west of Earth. STEREO-B was at a distance of 1.014~AU and an angular separation of $70^\circ$ east of Earth. 
An eruption (hereafter, CME1) was detected first by COR1-B at 16:05 UT where it appeared as a western limb eruption (from the spacecraft perspective). It was detected by COR1-A at 16:35 UT as a eastern limb eruption. Observations of CME1 in SDO, COR2-A and COR2-B  are shown in the top row of Figure~1.
The CME appears to be deflected slightly southward towards the ecliptic as it propagates through the COR1 FOV. It entered the FOV of COR2-A at 17:24 UT and of COR2-B at 17:54 UT. The radial speed in COR2-A was 362~km~s$^{-1}$ and the central position angle (PA) about 95$^\circ$. Speed and central PA in COR-2 quoted here and thereafter are from the CACTUS database \citep[]{Robbrecht:2004} unless noted otherwise, and the speeds are plane-of-the-sky speeds.  In COR2-B, the speed was  378~km~s$^{-1}$ and the central PA about 275$^\circ$. CME1 was also observed by LASCO/C2 starting at 18:30 UT as a faint asymmetric halo with a central PA of 320$^\circ$. The average speed of the halo CME was 278~km~s$^{-1}$ (values from the CDAW CME list). We determined the mass of this CME using the procedure of \citet{Colaninno:2009}. We found a mass of 1.5$ \pm 0.1 \times 10^{16}$~g. 

CME1 first entered the HI1-A FOV at 21:29 UT on May 23 and HI1-B at 22:09 UT. It had is a typical 3-part structure with a dense core that entered the HI1-A and HI1-B FOVs at 02:09 UT on May 24. The CME first entered  HI2-A at 22:09 UT on May 24 and HI2-B at 00:09 UT on May 25. \citet{Song:2012} recently reported and analyzed a series of blobs propagating and interacting in the corona within the current sheet behind CME1 from 04:00 to 13:00 UT on May 24, as well as an associated type-III radio burst detected by STEREO-A at around 10:24 UT on May 24.
There was also an unrelated narrow CME at around 06:00UT on May 24 observed at PA 66$^\circ$ in COR2-A and at PA 310$^\circ$ in COR2-B. 

\begin{figure*}[tb]
\begin{center}
{\includegraphics*[width=16.cm]{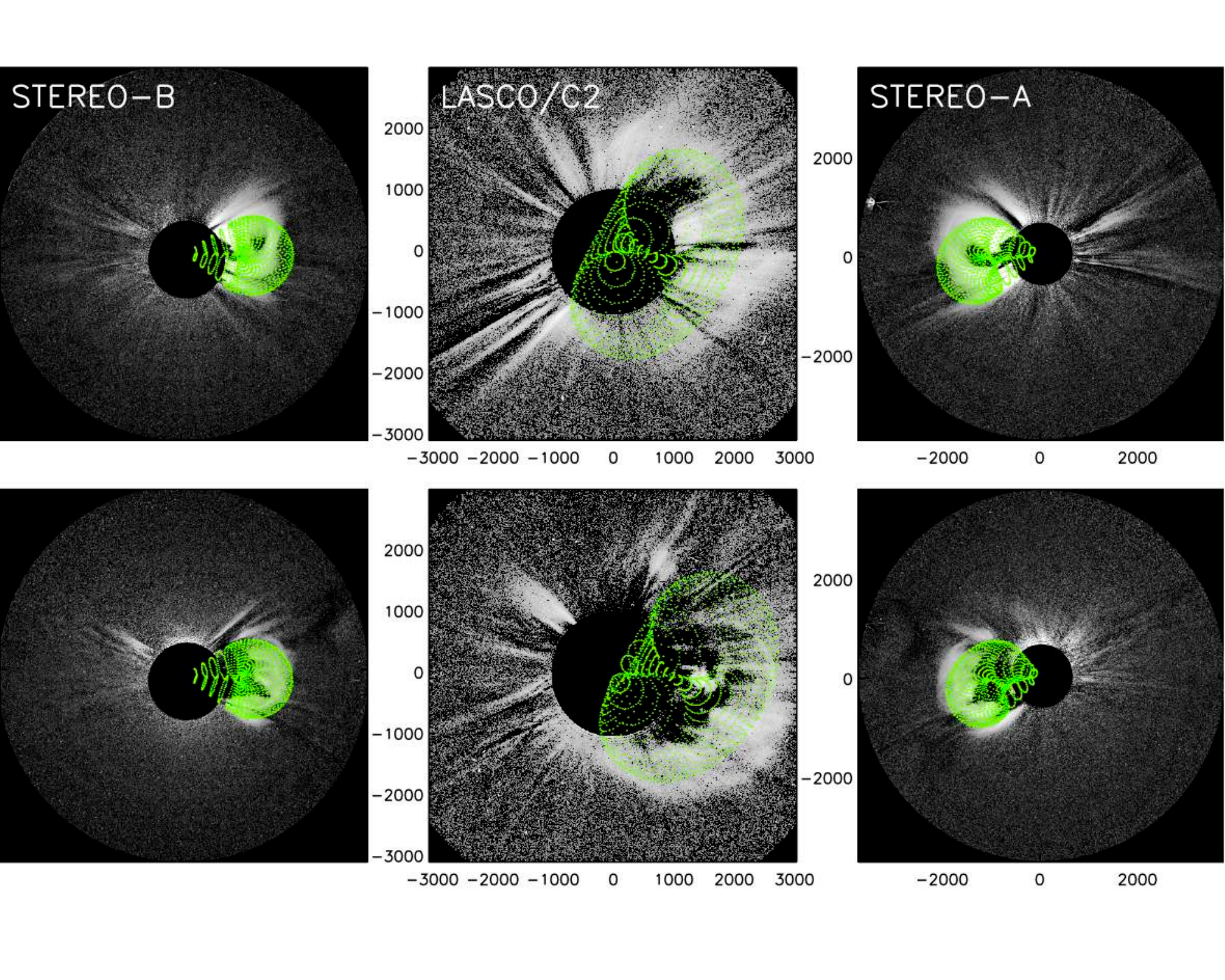}}
\caption{Graduated Cylindrical Shell (GCS) fitting of CME1 (top) and CME2 (bottom). The images are from STEREO-B/COR2, LASCO/C2 and STEREO-A/COR2, from left to right and the green overlay shows the fitted shape. See texts for details.}
\end{center}
\end{figure*}

\subsection{Second CME}

The second CME (hereafter, CME2) studied here was associated with a B1 flare and a filament eruption from N18W26 observed by SDO/AIA and SOHO/EIT. It was the same filament channel erupting as for the previous CME, but due to solar rotation it had moved by about 12$^\circ$ more to the west. 
An eruption was detected first by COR1-B at 13:05 UT on May 24, where it appeared as a western limb eruption. It was detected by COR1-A at 13:45 UT as an eastern limb eruption.  Observations of CME2 in SDO, COR2-A and COR2-B are shown in the bottom row of Figure~1.  
CME2 entered the FOV of COR2-A at 14:24 UT and the FOV of COR2-B at 14:54 UT. The speed in COR2-A was about 500~km~s$^{-1}$ and the central position angle (PA) about 95$^\circ$. In COR2-B, the speed was about 520~km~s$^{-1}$, and the central PA approximately 270$^\circ$. These speeds correspond to the ``core'' of the CME, which propagated to the south of the ecliptic (near PA 110$^\circ$, as seen from A and PA 250$^\circ$, as seen from B). In addition to this structure, which was relatively narrow (40$^\circ$ wide in PA), there was a much wider (90$^\circ$ wide) and faster ``leading edge'' centered around the ecliptic whose speed was around 650~km~s$^{-1}$. CME2 was also observed by LASCO/C2 starting at 14:06 UT on May 24 as a faint asymmetric halo with a central PA around 280$^\circ$. The average speed of the halo CME was 427~km~s$^{-1}$.
Using the procedure of \citet{Colaninno:2009}, we find a mass of 1.0$ \pm 0.1 \times 10^{16}$~g for CME2. 
CME2 entered the HI1-A FOV at 18:09 UT on May 24 and HI1-B at 17:29 UT. It entered the HI2-A FOV at 06:09 UT on May 25 during a data gap of STEREO-B/SECCHI.

\subsection{Indications of a Probable Interaction}
Assuming constant speeds of 380~km~s$^{-1}$ and 500~km~s$^{-1}$ (as reported on the CACTUS CME catalog) for the two CMEs, the two leading edges should cross at around 0.8~AU. Assuming speeds of 400~km~s$^{-1}$ and 650~km~s$^{-1}$ (speed of the fastest fronts), the interaction should have happened by 0.55~AU. While a deceleration of the CMEs is expected, interaction between the front of CME2 and the back of CME1 is anticipated to happen much closer to the Sun than the distance estimated above. We also note that the two CME fronts overlap in the HI2 FOV, possibly indicating interaction of the CMEs. Background-subtracted images of the two CMEs in the HI1 FOV are shown in Figure~2, corresponding to 2010 May 25 at 00:09 UT (top row) and 06:49 UT (bottom row).

\begin{figure*}[tb]
\begin{center}
{\includegraphics*[width=8.4cm]{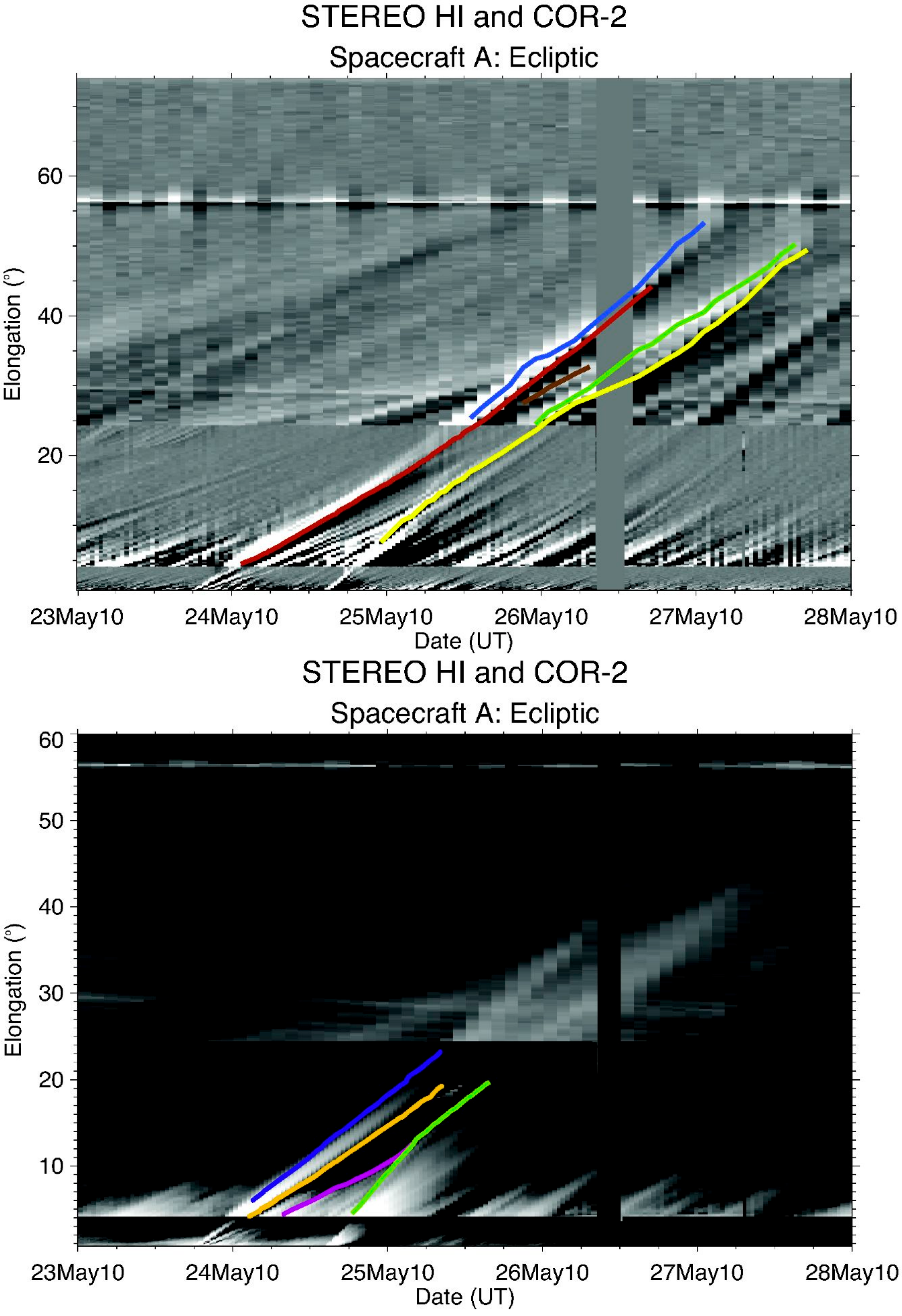}}
{\includegraphics*[width=8.4cm]{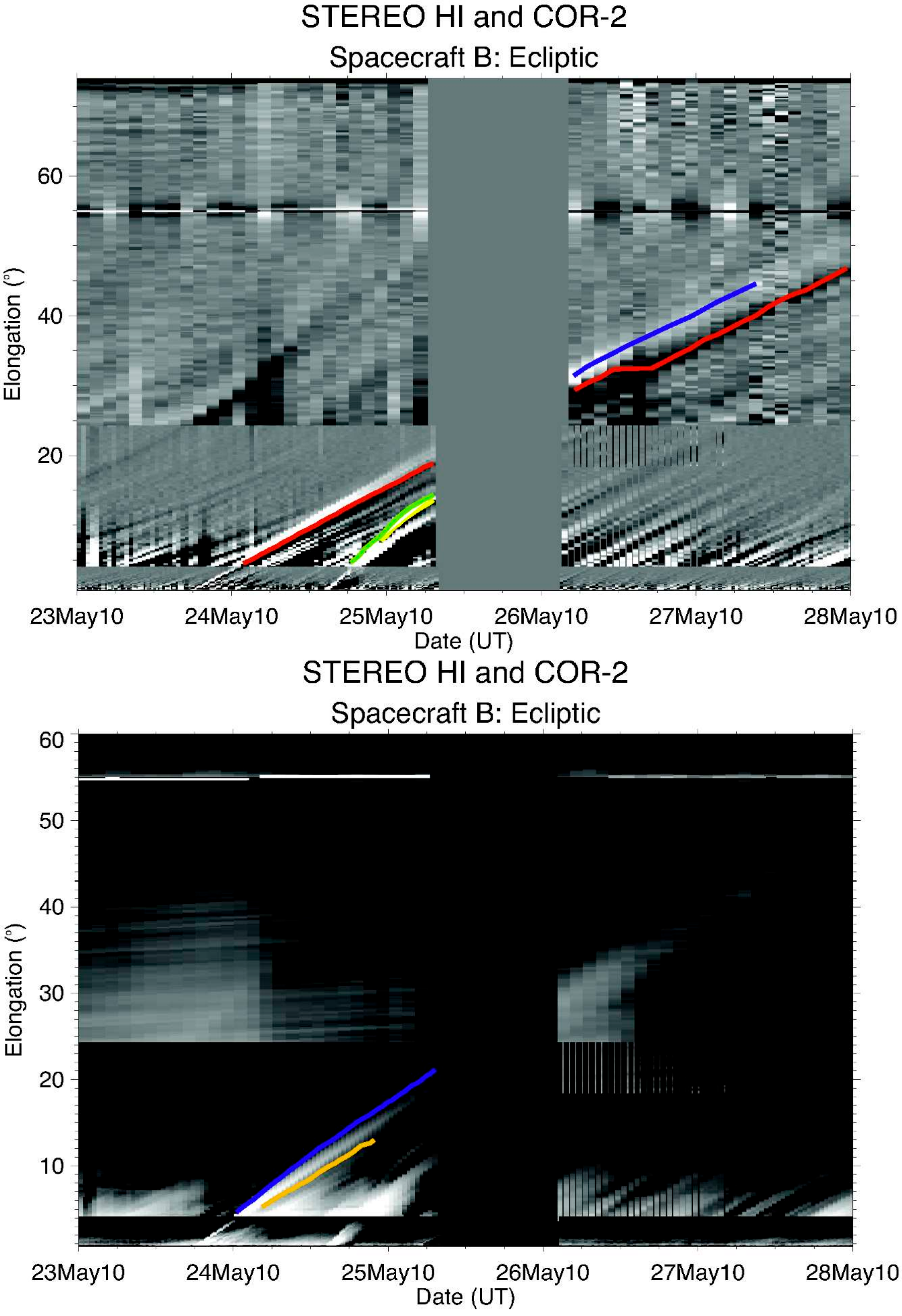}}
\caption{J-maps for the two CMEs and the discussed tracks. Left: STEREO-A, right: STEREO-B. Top: Running-difference, bottom: background-subtracted. From left to right, there is the leading edge of CME (shown in blue), its black edge (BE1, red), the end of the leading edge or beginning of the cavity (orange), the end of the CME1 (pink), the leading edge (green) and black edge (yellow) of CME2 (in red) as shown in the J-maps. Brown is used to highlight a secondary front in HI2-A discussed in the text.}
\end{center}
\end{figure*}

\section{DIRECTION OF THE BRIGHT FRONTS} \label{direction}

Determining the directions of the bright fronts observed by SECCHI is essential  for understanding the kinematics and evolution of the two CMEs. Southward deflection away from the flare site was observed for both CMEs. Therefore, we cannot simply use the flare location as the direction of propagation of the CMEs. Single-spacecraft fitting methods \citep[]{Sheeley:1999, Rouillard:2008, Lugaz:2010c,Davies:2012} can only be used if the speeds of the CMEs are relatively constant. In the case of a CME-CME interaction, there would be  large variations in the CME speeds. Finally, it is not even clear which bright front in HI2 corresponds to which CME.

For these reasons, we rely on stereoscopic methods: triangulation \citep[see][]{Liu:2010a}, which is based on the Fixed-$\Phi$ (F$\Phi$) geometry; and tangent-to-a-sphere method \citep[see][]{Lugaz:2010b}, which is based on the Harmonic Mean (HM) geometry. Triangulation has been mostly used in the EUVI, COR1, COR2 and HI1 FOVs, and it has been shown \citep[e.g., see][]{deKoning:2009} to yield reliable results as compared to other methods. Because it assumes that the same plasma element is being observed by both spacecraft, triangulation is expected to break down at large elongation angles and for wide CMEs. The tangent-to-a-sphere method takes into account the fact that the same part of the CME is not imaged by the two spacecraft. It is expected to yield better results for those CMEs observed as a halo by one of the spacecraft, or for which the width cannot be neglected. We also use the visual fitting of the CMEs in the coronagraphs FOV using the Graduated Cylindrical Shell (GCS) model \citep[]{Thernisien:2009, Thernisien:2011}, which assumes a flux rope shape. In the next section, we analyze the observations in the COR2 and the HI1 FOVs to derive the CME directions. 

\subsection{Direction of the CMEs Using the GCS Model}

We use the three coronagraphic views on CME1 at around 20:30 UT on May 23 from STEREO-A/COR2 (at 20:24 UT), STEREO-B/COR2 (at 20:24 UT) and LASCO/C2 (at 20:30 UT). The three views are shown in the top row of Figure~3. At this time, the CME was at a distance of around 9.5~$R_\odot$. The best visual fit for CME1 is a direction of S0W10 with a large tilt angle of  65$^\circ$  with respect to the ecliptic, and a small half-angle of 15$^\circ$. For a discussion of the uncertainties of the method, please refer to \citet{Thernisien:2009}. Typical uncertainties are in the range of $\pm 5^\circ$, $\pm 10^\circ$, and  $\pm 20^\circ$ for the longitude, the half-angle, and the tilt, respectively.

For CME2, we use the three views at around 16:30 UT on May 24, at which time CME2 was also around 9.5~$R_\odot$ (the three views are shown in the bottom row of Figure~3, corresponding to 16:24 UT for the COR2 views and 16:30 UT for the LASCO view). The best fit for CME2 is a direction of S02W26, with a tilt angle of 60$^\circ$ and a half-angle of $20^\circ$. As can be seen from Figure~3, for both CMEs, the LASCO view is essential for constraining the direction of the CMEs since both CMEs had a similar aspect in STEREO-A and STEREO-B.

\begin{figure}[ht*]
\begin{center}
{\includegraphics*[width=8.3cm]{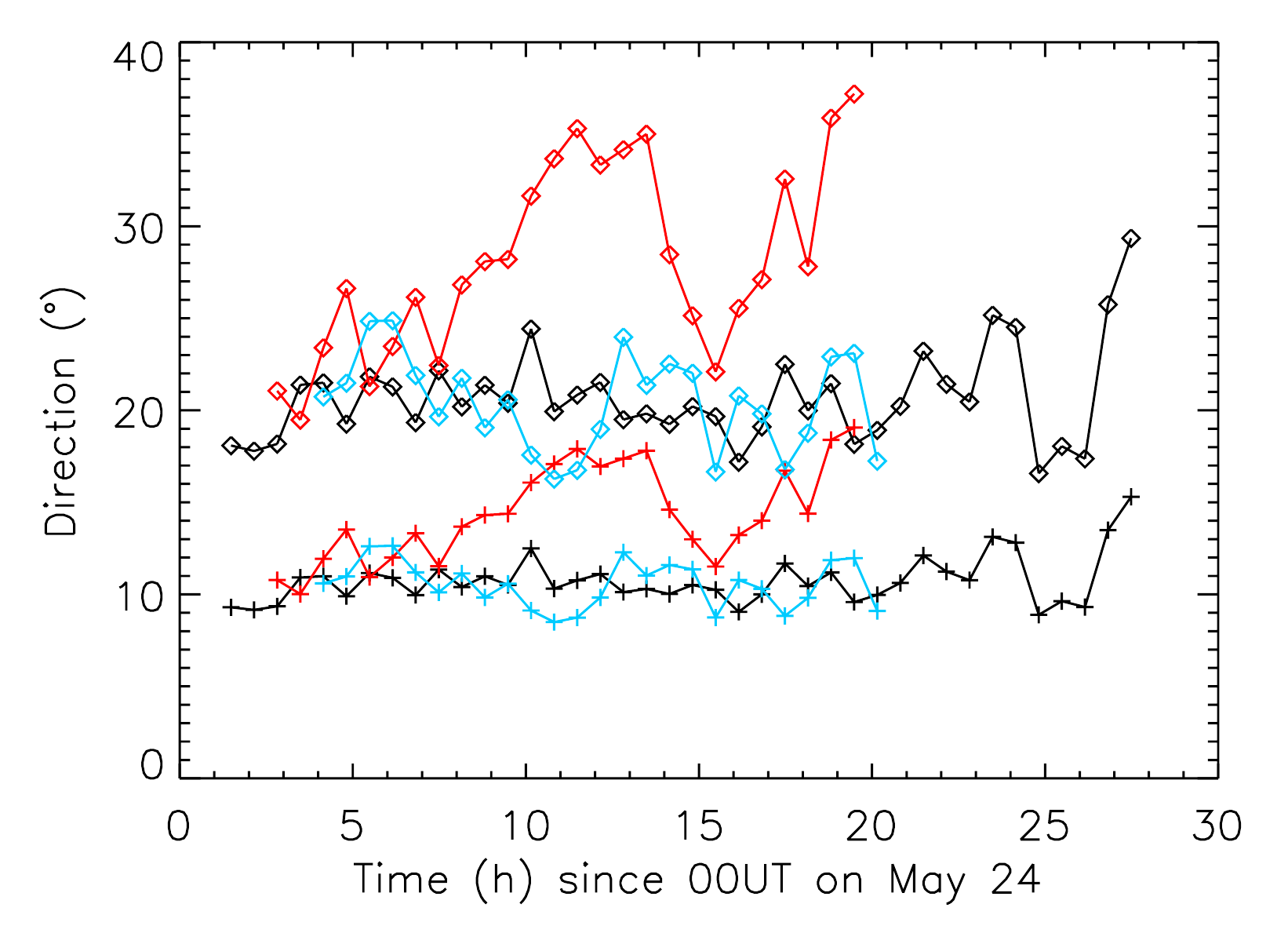}}\\
{\includegraphics*[width=8.3cm]{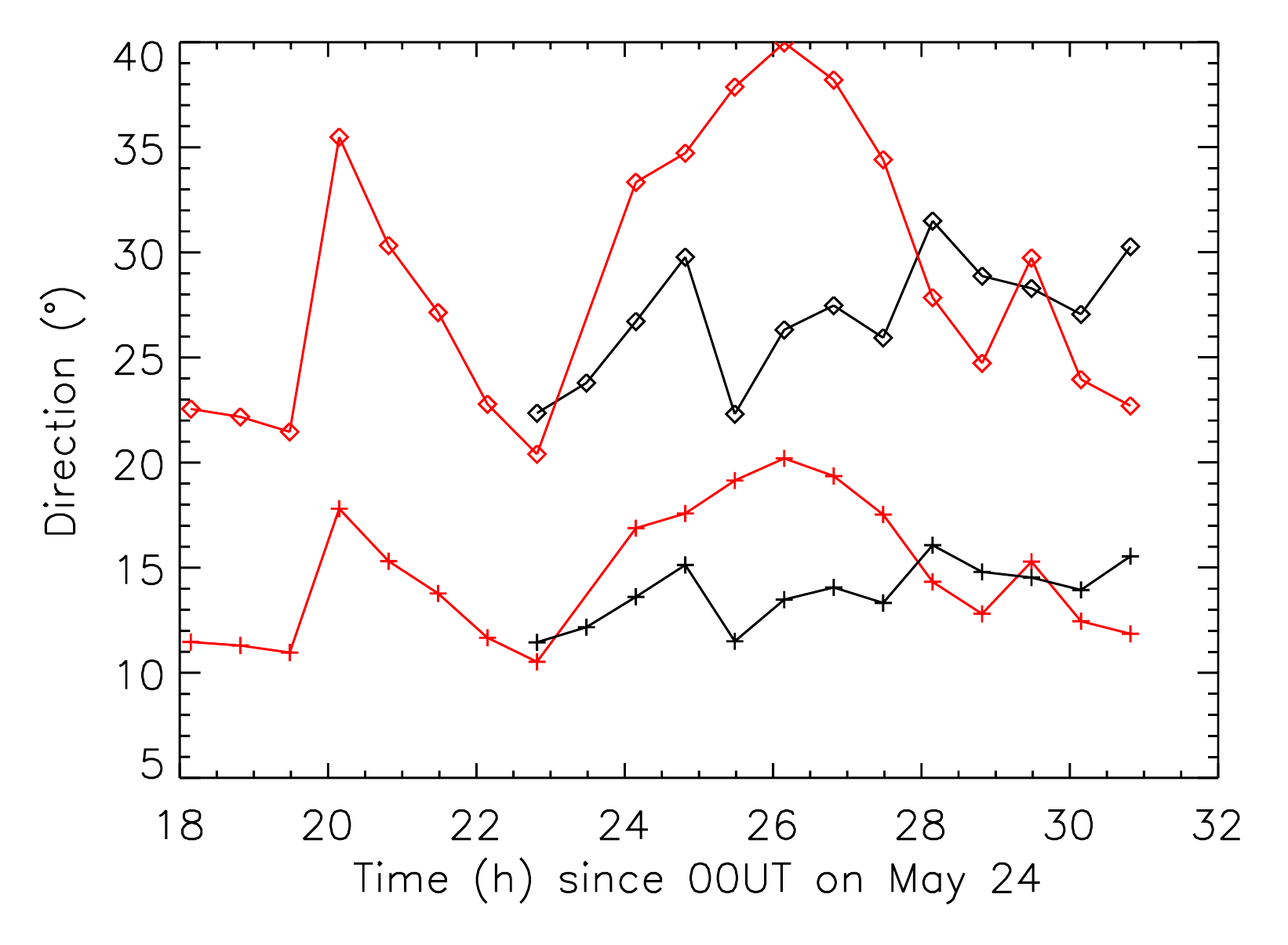}}\\
\vspace{0.2cm}
{\includegraphics*[width=8.3cm]{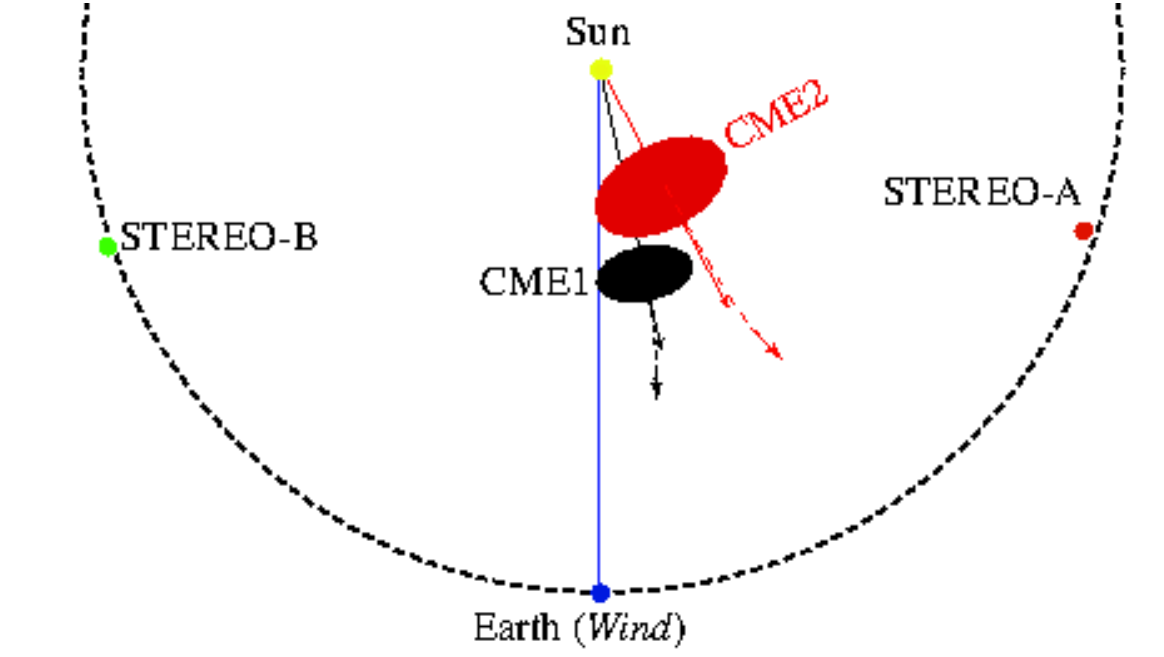}}
\caption{Direction of the first (top) and second (middle) CMEs. Crosses are used for results from the triangulation method and diamonds for the tangent-to-a-sphere method. The leading edges are shown in red, the ``black'' edges in black and the back of the CME in blue. The bottom panel shows a sketch of the geometry of the observations on 2010 May 24-27 as seen from the ecliptic north. The initial directions (12$^\circ$ and 28$^\circ$) are shown with solid arrows and the deflection during the interaction is shown with dashed arrows (see text for details).} 
\end{center}
\end{figure}

\subsection{Direction of CME1 in HI1 FOV}

In the HI1 FOV, we first track three features associated with CME1: the leading edge (as seen in background-subtracted or running-difference J-maps); the ``black'' edge (the transition from white to black, see below) as seen in running-difference J-maps; and the back of the leading edge as observed in background-subtracted J-maps. J-maps are time-elongation maps created following the procedure explained in \citet{Sheeley:1999} and \citet{Davies:2009}.  In the bottom panels of Figure~4 (background-subtracted J-maps), the leading edge of CME1 is highlighted in blue and the back of the leading edge in orange. In the top panels of Figure~4 (running-difference J-maps), the ``black'' edge is highlighted in red.

The ``black'' edge corresponds to the apparent end of the leading edge as seen in the running-difference images. This is easier to track because the contrast is optimal between bright and dark regions as compared to the real leading edge, which is between grey and bright regions. 
However, 
it is not clear to what physical part of the CME front (if any) the ``black'' edge corresponds. By tracking three different features, we attempt to determine if all parts of a CME can be reasonably assumed to have the same direction. 

Using triangulation, the propagation direction of the leading edge is derived to be about 14$^\circ$ (positive angles refer to west of the Sun-Earth line), the direction of the ``black'' edge is about 10.5$^\circ$ and that of the back of the CME is about 10.5$^\circ$. In all cases, the direction remains almost constant with distance until about 14-15$^\circ$ elongation when it increases. 
Using the method of \citet{Lugaz:2010b}, the CME direction is found to be 27.5$^\circ$, 20.5$^\circ$ and 20.5$^\circ$ for the leading edge, the ``black'' edge, and the back of the CME, respectively. Results using HI1 data are summarized in the top panel of Figure~5.

This Figure shows two important results. 1) The direction as derived using the actual leading edge, the ``black'' edge, or the back of the CME are within the typical error bar of the methods ($\pm 5^\circ$), with the direction obtained from the ``black'' edge being more steady than that derived from the actual leading edge. This validates all previous studies which have typically used the black edge of the CME to derive its direction and kinematics. 2) Our study shows that, for the first 24 hours of its propagation in the heliosphere (past 20~$R_\odot$), the CME has a constant direction, independently of the method used to derive this direction. Finally, it should be noted that 3) the CME heliocentric distance in the HI1 FOV has only a very weak dependence on the method chosen to derive the propagation direction. This has been pointed out before, especially for fitting methods;  both the HM and F$\Phi$ fitting tend to produce different directions and speeds, but similar arrival times (for example, see \citet{THoward:2009a}, \citet{Moestl:2011}, \citet{Lugaz:2012b}). 

We believe that the direction of CME1 derived by  triangulation should be trusted because: i) it corresponds to the approximate location of the source region of the CME (W12) as well as the direction obtained from the GCS fitting, and, ii) the CME is relatively narrow and the assumptions of the method of \citet{Lugaz:2010b} are not valid in this case. 
For the rest of the analysis, we use a direction of 12$^\circ$ (the average of the direction from triangulation of the three tracked features) and we use the F$\Phi$ approximation in HI1 to determine the distance and kinematics for all tracks associated with CME1.

\subsection{Direction of CME2 in HI1 FOV}

We follow the same procedure as described above for CME2. It should be noted that the edges of CME2 are less well defined because of the presence of bright features associated with CME1. In the bottom panels of Figure~4, the leading edge of CME2 is highlighted in green and, in the top panels of Figure~4, the ``black'' edge of CME2 is highlighted in yellow.
Using triangulation, the direction of the leading edge is about 15$^\circ$ and the direction of the ``black'' edge is about 14$^\circ$. Using the method of \citet{Lugaz:2010b}, the CME direction is 29$^\circ$ and 27$^\circ$ for the leading and the ``black'' edges, respectively. 

We believe that the direction of the CME derived by the tangent-to-a-sphere method should be used, because it corresponds to the approximate location of the source region of the CME (W26) as well as the direction obtained from the GCS fitting (W26). Hereafter, we use the HM approximation to analyze observations of CME2 in HI1. A sketch of the geometry including the two CMEs, as seen from ecliptic north, is shown in the bottom panel of Figure~5.

\section{EVOLUTION OF THE CMES IN HI1 AND HI2 FOV} \label{HI1}

\subsection{HI Data}

In addition to the leading edges (real and black, as discussed above), we track the back of the leading edge of CME1 as seen in background-subtracted images (orange track in Figure~4). At 1~AU, magnetic ejecta typically have low density \citep[]{Burlaga:1981}. Magnetic flux ropes are thought to be present in the dark cavities of three-part CME structures \citep[e.g., see][]{Hundhausen:1993}. Ejected high-density filament material (the CME core) is at first embedded inside this cavity and, at later times, it is expected to over-expand and become part of the low-density region. Therefore, the back of the leading edge can be assumed to correspond approximately to the front of the magnetic ejecta. Recently, \citet{THoward:2012} have shown how a magnetic cloud observed at Earth can be traced back to the dark cavity as seen in background-subtracted images in coronagraphs and heliospheric imagers. Density enhancements or V-shape structures are often observed by SMEI and the HIs corresponding to the back of the magnetic ejecta \citep[]{Kahler:2007, Harrison:2008}. We also track one such feature at the back of CME1 (pink track in Figure~4), which corresponds to the compression at the back of the cavity or magnetic ejecta \citep[]{THoward:2012}. 
The measurements of the five tracked fronts in HI1 are summarized in the top left panel of Figure~6.


We find that the temporal offset between the leading edge (asterisks) and the black edge (triangles) increases from 2 hours to about 3.3 hours as CME1 propagates in the HI1-A FOV. In the HI2-A FOV, we track the apparent brightness maximum (or center of the track) in the running-difference J-map as well as the black edge, and we find a temporal offset between the two tracks of about 3 hours. Assuming that the apparent center of the track is the actual center of the leading edge, this means that the temporal offset between the leading edge and the black edge of CME1 is about 6 hours. An offset of about 4-6 hours was also estimated in \citet{Lugaz:2012b}  and, in general, it should be corrected in studies of CME arrival times using the black track. 
The offset is approximately equal in HI1-B to what is found in  HI1-A. 

This finding means that the black edge seen in running-difference images and J-maps is not simply a result of the running-difference procedure, since the offset between the CME leading edge and the black edge is not constant within an instrument. The increasing offset is most likely due to the expansion of the CME sheath. However, there is a strong difference in the value of the offset between HI1 ($\sim 2-3$ hours) and HI2 ($\sim 4-6$ hours), which indicates that this offset is also partially controlled by the resolution and frequency of the observations. 

\begin{figure*}[ht*]
\begin{center}
{\includegraphics*[width=8.4cm]{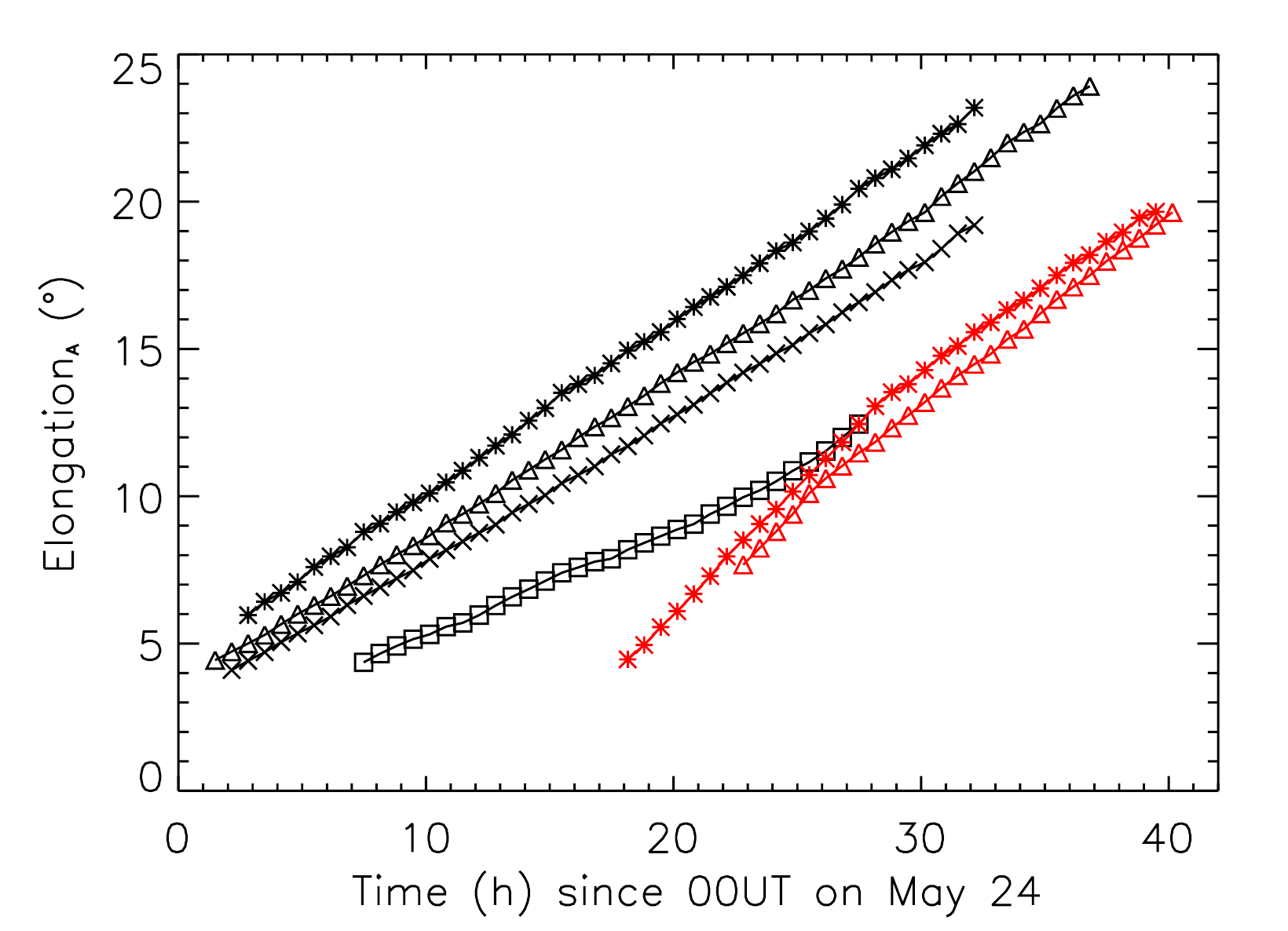}}
{\includegraphics*[width=8.4cm]{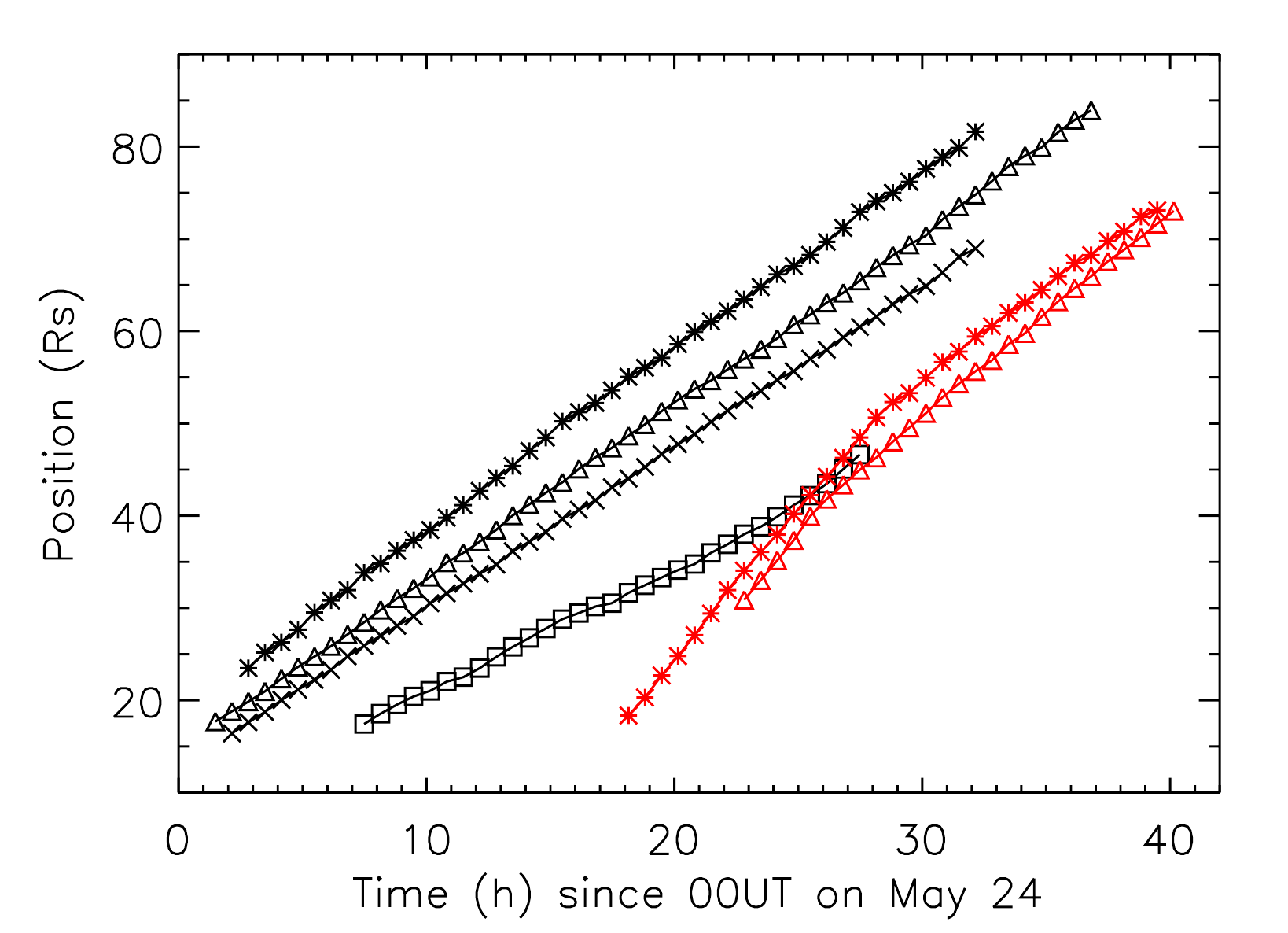}}\\
{\includegraphics*[width=8.4cm]{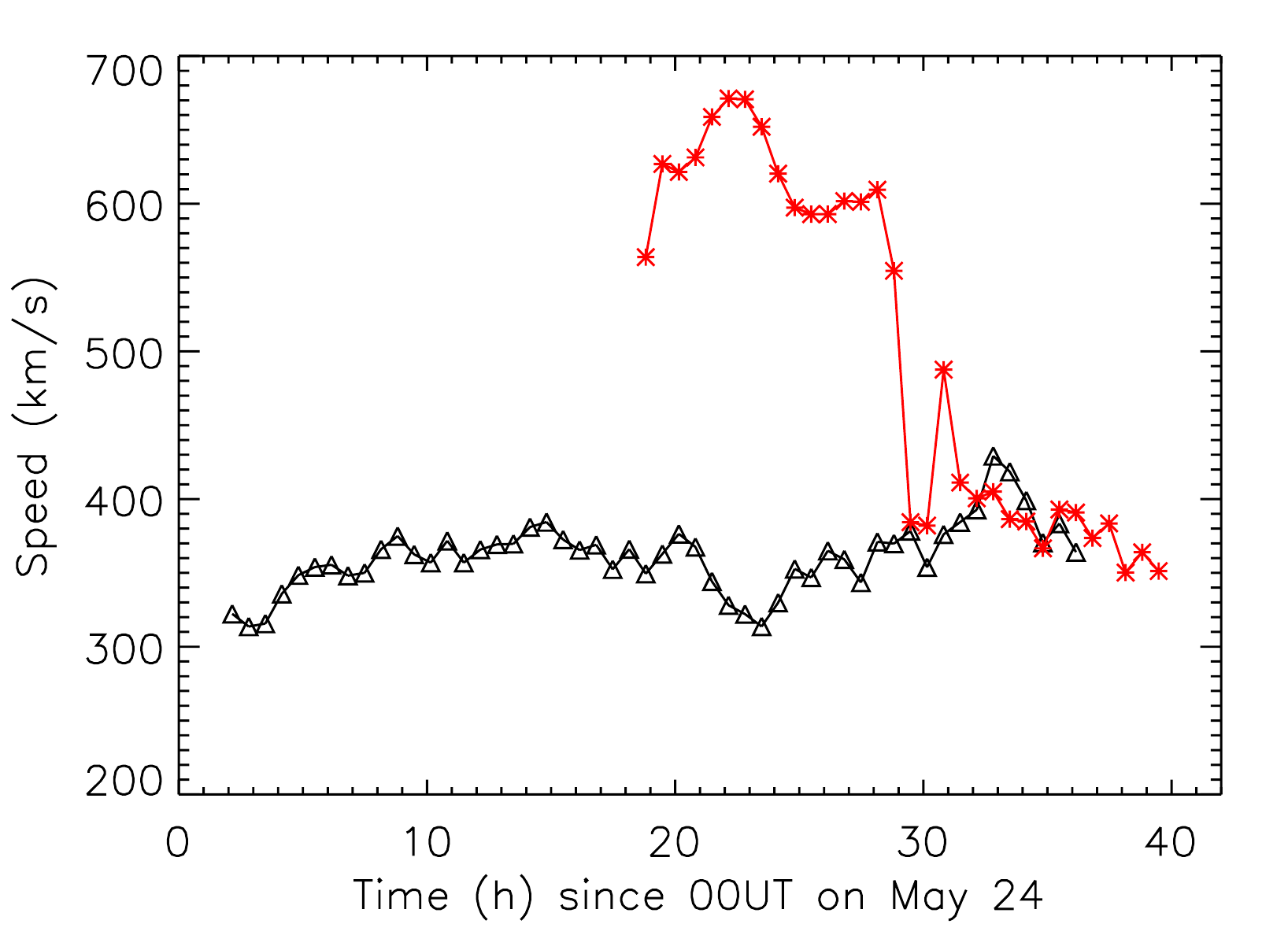}}
{\includegraphics*[width=8.4cm]{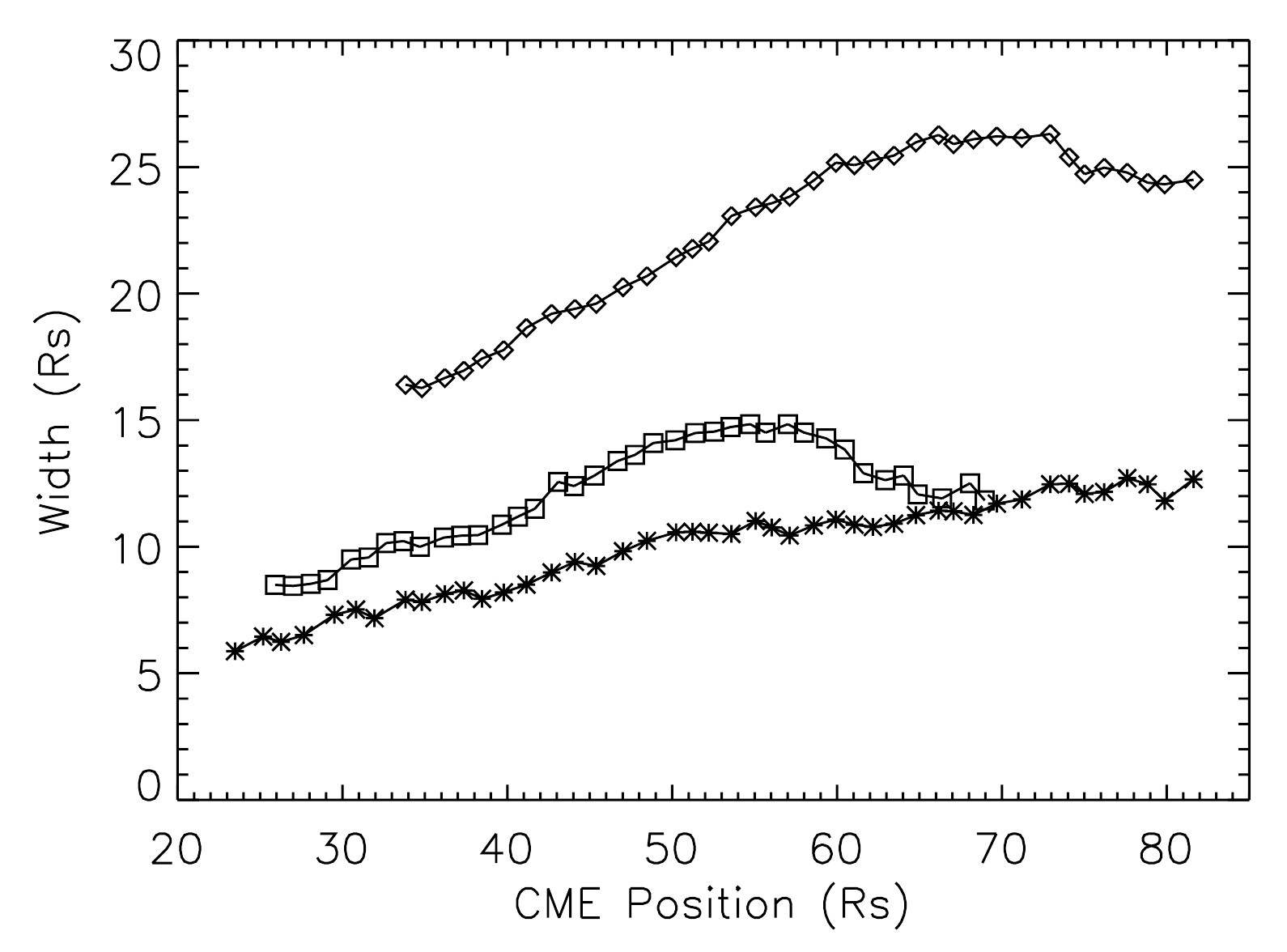}}
\caption{Top left: Time-elongation measurements in HI1-A of the tracked fronts, corresponding to the leading edge of CME1, its black edge, the end of the leading edge or beginning of the cavity, the end of the CME1, all in black, and the leading edge and black edge of CME2 (in red) as shown in the J-maps. Top right: Heliocentric distances of the tracked fronts for CME1 (black) and CME2 (red). The fronts are the same as described in top left with the same conventions. Bottom left: Speed of the black edge of CME1 and leading edge of CME2. Bottom right: Widths of CME1; diamonds: total ICME (sheath + ejecta), stars: sheath, and square: ejecta. Error estimates are $\pm 0.6~R_\odot$ for the distances, $\pm 1.2~R_\odot$ for the width, and $\pm 85$~km~s$^{-1}$ for the velocities (see text for details).}
\end{center}
\end{figure*}

\subsection{Determining the distances of the CMEs in the inner heliosphere}

In the top right panel of Figure 6, we show the time-distance evolution of the different tracks. As done in \citet{Lugaz:2010b}, we determine the typical error  associated with the manual selection of the front by estimating the accuracy of the elongation measurements (as shown in the top left panel of Figure~6) to be $\pm 0.15^\circ$ (only valid for HI1 data). Then, the error on the distances is typically $\pm 0.6~R_\odot$ and the error on the velocity is $\pm 85$~km~s$^{-1}$. 
In the bottom left panel of Figure~6, we show the speed of the leading edge of the two CMEs; we use the black edge of CME1, because the results are steadier, and the leading edge of CME2 because the black edge is not well defined early on (see the yellow track in Figure~4).

\subsection{Width of the First CME}

Using the data obtained from HI1-A, we calculate some of the radial widths (or radial extents) associated with CME1: total ICME width, total ejecta width (without the sheath), and sheath width. Based on the typical error for the distance, the widths have an uncertainty of $\pm 1.2~R_\odot$. A similar study was recently completed by \citet{Savani:2012} for four CMEs observed by SECCHI. 
As can be seen in the bottom right panel of Figure~6, the three widths increase with distance, except that the ejecta and total ICME width become constant or start to decrease on May 25 at 00:09UT. This does not appear to be connected with the front of the CME,  as is clear from the fact that the width of the sheath region continues to increase. This marks the probable beginning of the interaction between the two CMEs. 

We fit the widths, $W$ as power laws with respect to the radial distance, $r$ before the start of the CME-CME interaction.  We find the following relationships:
$W_\mathrm{ICME} = 0.30 r_\mathrm{ICME}^{0.76}$, $W_\mathrm{ejecta} = 0.21 r_\mathrm{ejecta}^{0.82}$ and $W_\mathrm{sheath} = 0.11 r_\mathrm{ICME}^{0.54}$, where $W$ and $r$ are both expressed in AU.
The value of the power index for the width of the ejecta and the ICME are very close to the relation of \citet{Bothmer:1994}, who found a power law of 0.78 based on {\it in situ} measurements at 1~AU for magnetic clouds, and relatively close to the power-law values of 0.92 and 0.91 found by \citet{Liu:2005} and \citet{Gulisano:2010}, respectively. All these studies also find a typical value for the ejecta width at 1 AU of 0.24$\pm 0.02$~AU, close to the value of 0.21~AU found here.  Here, contrary to many other studies, the sheath is not made of shock-compressed solar wind, because  CME1 is probably too slow to drive a shock. What is referred to as the sheath is a combination of swept-up solar wind mass and possibly some coronal mass (part of what initially comprises CME1).

Finally, we make a crude estimate of the angular half-width of CME1 in ecliptic longitude. To do this, we assume that the radial width as derived here is equal to the CME longitudinal width. This assumption is not arbitrary, since, in fact, the axis of CME1 is found to be highly tilted using the GCS fitting (see also measurements at 1~AU) and CME cross-sections close to the Sun are expected to be nearly circular. Under these assumptions, we find that the angular half-width of CME1 is about 8--10$^\circ$. A different, and more rigorous, estimate can be obtained using the stereoscopic method (model 2) of \citet{Lugaz:2010b}, which allows to derive the actual CME1 longitudinal half-width under a set of assumptions. Using HI1 data and a fixed direction of $12^\circ$, we find that the half-width is $11^\circ \pm 2^\circ$, in relatively good agreement with what we derived above. Such a small half-width is consistent with the use of the F$\Phi$ approximation for this CME and has some important consequences when we consider the physical processes during the CME-CME interaction (section 6).

\subsection{Kinematics of the CMEs in HI1 FOV}

Initially on May 24, the leading edge of CME1 propagates with a speed of about 360~km~s$^{-1}$ and the leading edge of CME2 with a speed of 620~km~s$^{-1}$, corresponding approximately to the speed derived in COR2. A direct consequence of the difference in speeds is that CME2 quickly catches up with CME1. In addition to the speeds plotted in Figure~6, we find that the beginning of the magnetic ejecta (cavity) of CME1 has a speed of 340~km~s$^{-1}$ and that the end of CME1 has a speed of about 260~km~s$^{-1}$, reflecting the expansion of the magnetic ejecta (as described in the previous section). 

At 00:09 UT on May 25, CME1 starts to contract in radial extent, with the front and rear of CME1 becoming closer together, and at around this time (01:19 UT on May 25), the leading edge of CME2 decelerates to reach a near constant value of 380~km~s$^{-1}$ at 02:40 UT. This supports our claim that, indeed, the leading edge of  CME2 has collided with the back of the magnetic ejecta of CME1.
The material that makes up the leading edge of CME2 cannot propagate through the magnetic barrier and it is forced to propagate with the same speed as CME1. New material is not expected to accumulate between the two CMEs but the existing material cannot easily be removed until the two CMEs separate, if they do. The speed of this dense material is now constrained by the speed of CME1. 

\begin{figure*}[tb*]
\begin{center}
{\includegraphics*[width=5.7cm]{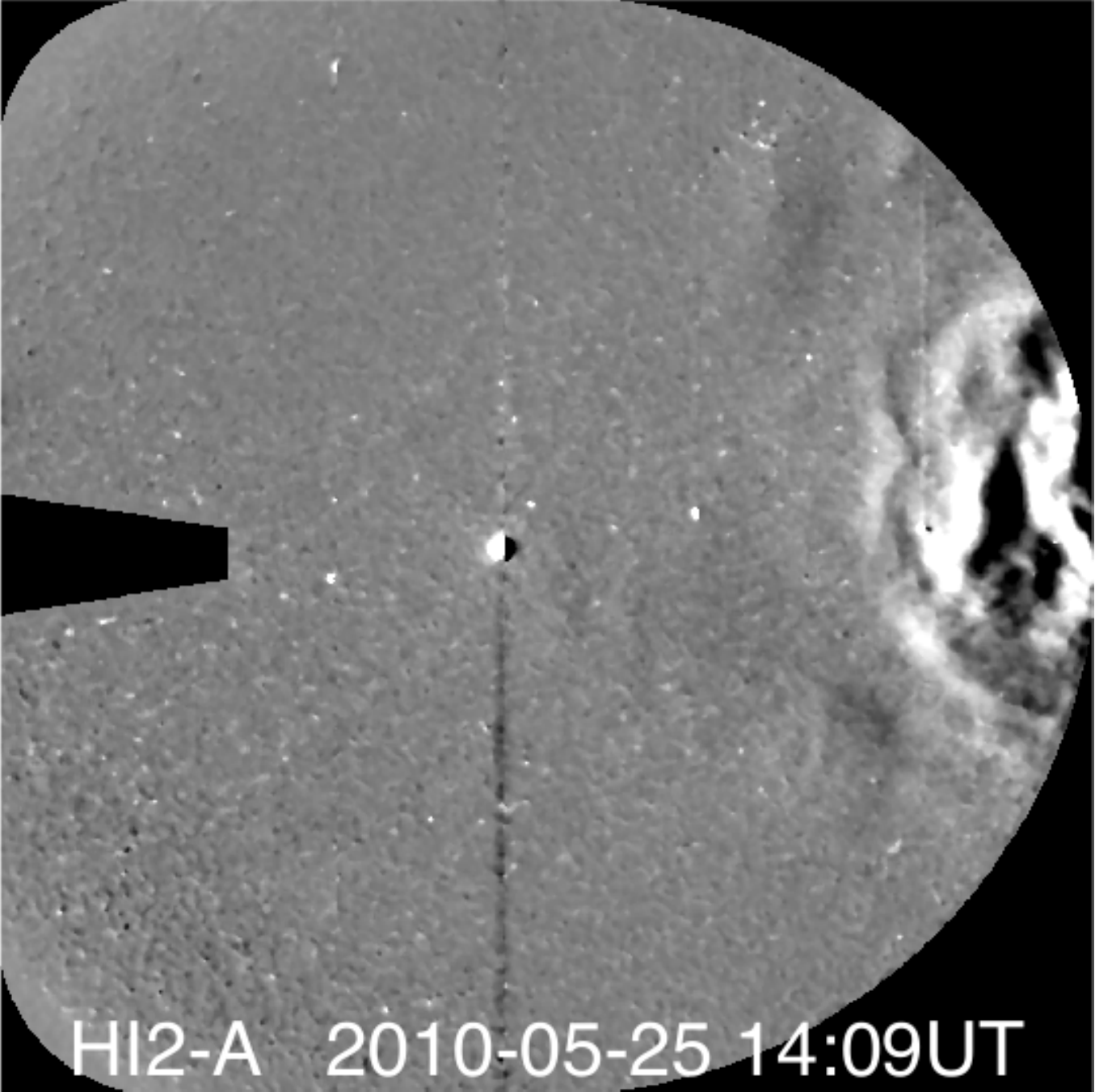}}
{\includegraphics*[width=5.7cm]{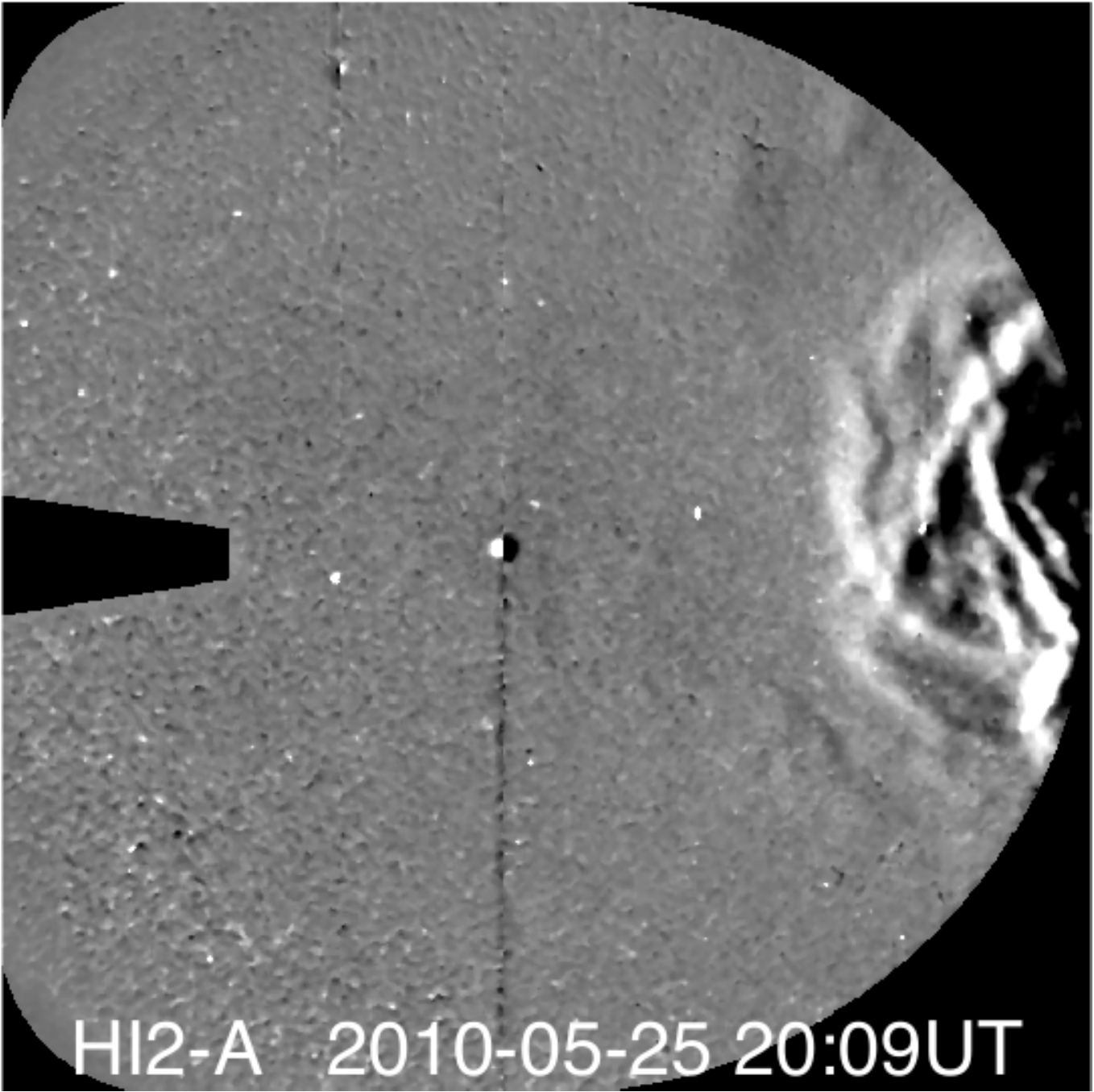}}
{\includegraphics*[width=5.7cm]{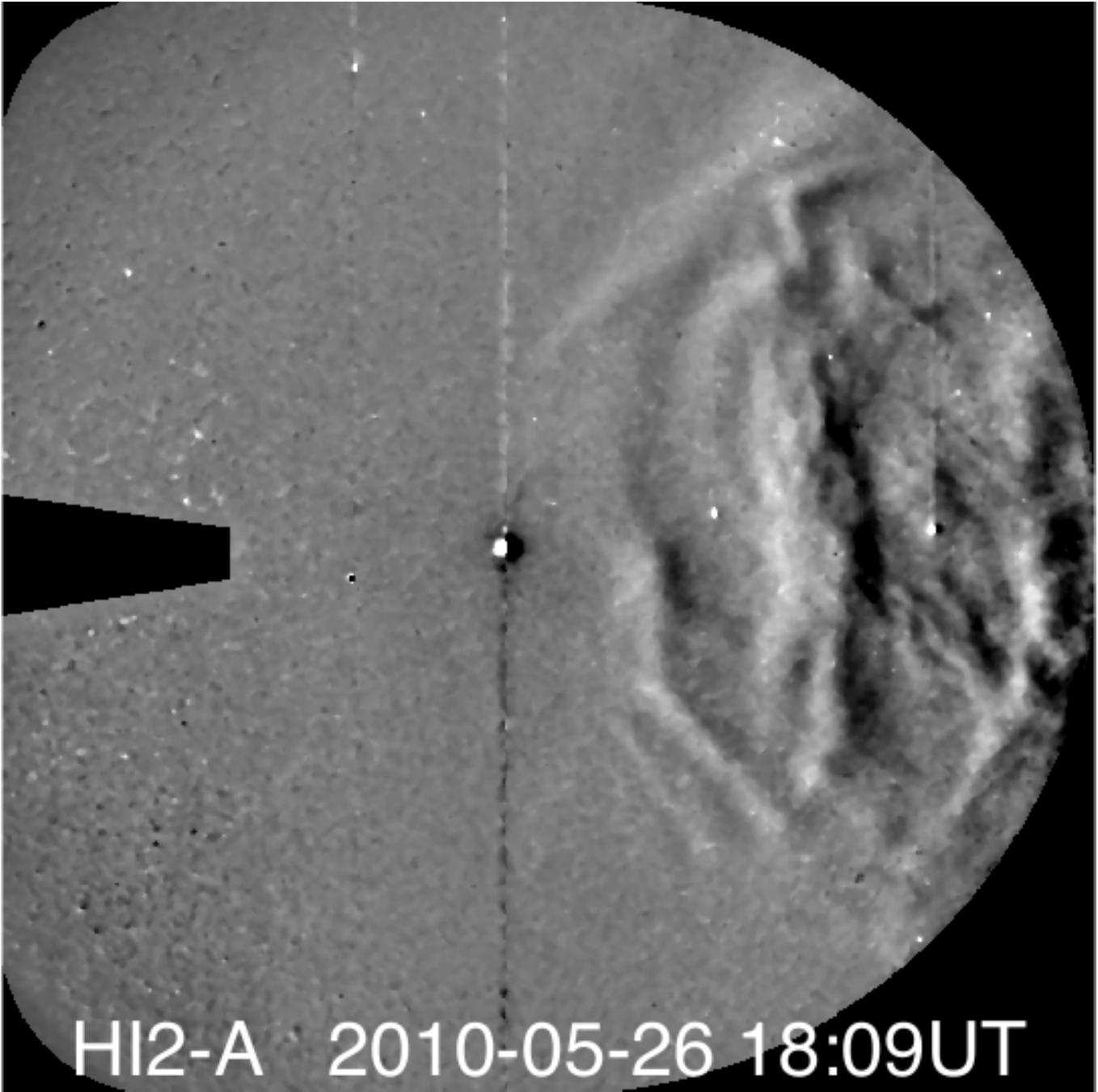}}
\caption{STEREO-A HI2 running-difference images on May 25 showing the ``splitting'' of the first front and its ``interaction'' with the second front and also illustrating the fact that it appears that after on May 26, CME2 is now ahead of CME1.}
\end{center}
\end{figure*}

\subsection{Evolution of the CMEs in the HI2 FOV} 

The change in temporal and spatial resolution as well as intensity threshold and the difficulty to use background-subtracted images, make it complicated to follow the evolution of the CMEs in the HI2 FOV. In addition, the long data gap in HI-B further muddles the picture. 
From a visual inspection of the J-maps, the first striking feature in HI2 is the fact that there are two bright tracks associated with the CMEs in HI2-A but only one bright track in HI2-B. Below, we quickly discuss the origin of these tracks.


The idea that the first and second tracks in HI2-A correspond to different CMEs before and after the short data gap in HI-A on May 26 can be inferred by inspecting the actual images (instead of the J-map). Throughout its propagation, CME1 appears relatively symmetric in position angle with respect to the ecliptic, whereas CME2 has an ear-shape with a flattening or change in indentation around position angle (PA) 110$^\circ$ (see Figure~2, for example). As shown in Figure~7, on May 26, the first CME observed in HI2-A has the shape characteristics of CME2. 

We want to emphasize the following: while running-difference images in HI2-A show two (or three) bright tracks at all times, separated by dark regions, until the beginning of May 26, there is in fact only one wide bright front as seen in background-subtracted images (as can be seen on the bottom left panel of Figure~4). In HI2-A at 14:09 UT on May 25, there is a ``split'' of the first track visible at around 25$^\circ$ (see left and middle panels of Figure~7). It appears as if  part of the first track becomes part of the second track 
(this transient third track is highlighted in brown in the running-difference J-maps). 
Finally, it is worth noting that the first track in HI2-A clearly has some geometrical acceleration (similar to what has been discussed in \citet{Rouillard:2008} for example). This, combined with the fact that no similar front is observed in STEREO-B images, shows that this front propagates in a direction away from the Sun-Earth line, more towards STEREO-A, as CME2 is sketched in the bottom panel of Figure~5. 

\begin{figure*}[ht*]
\begin{minipage}[b]{0.5\linewidth}
\centering
{\includegraphics*[height= 14cm]{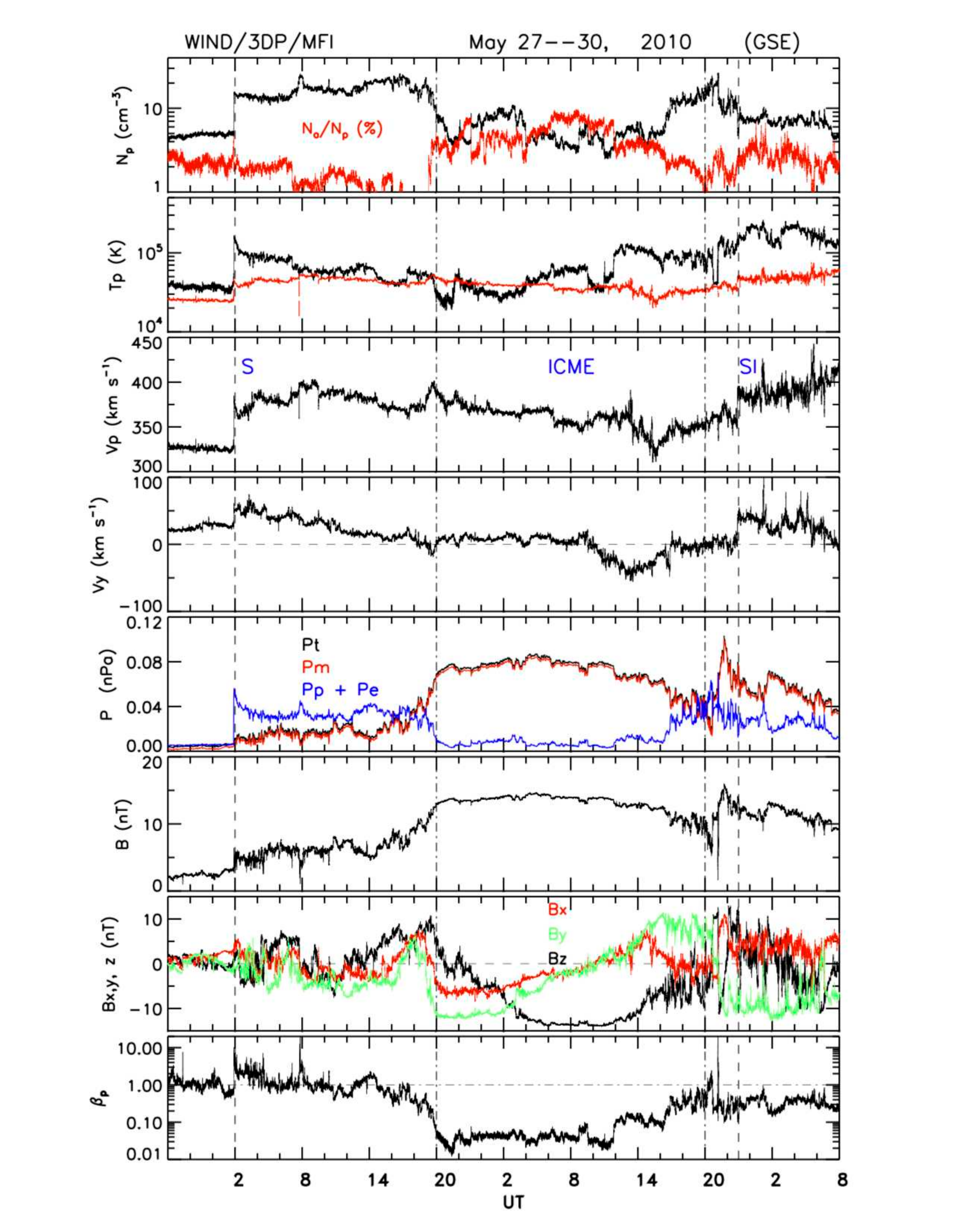}}
\end{minipage}
\begin{minipage}[b]{0.5\linewidth}
\centering
{\includegraphics*[height= 7cm]{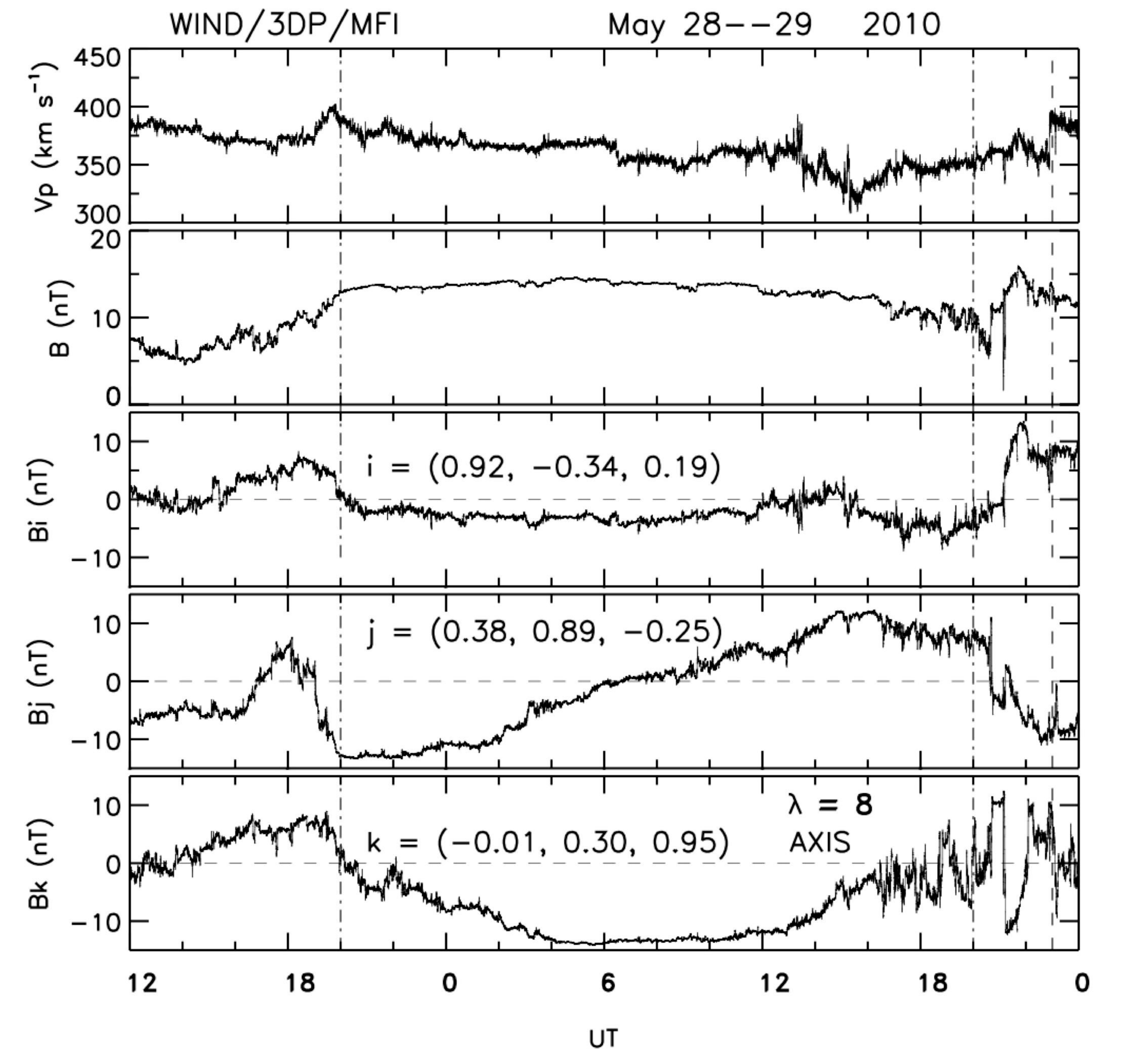}}
{\includegraphics*[width=7cm]{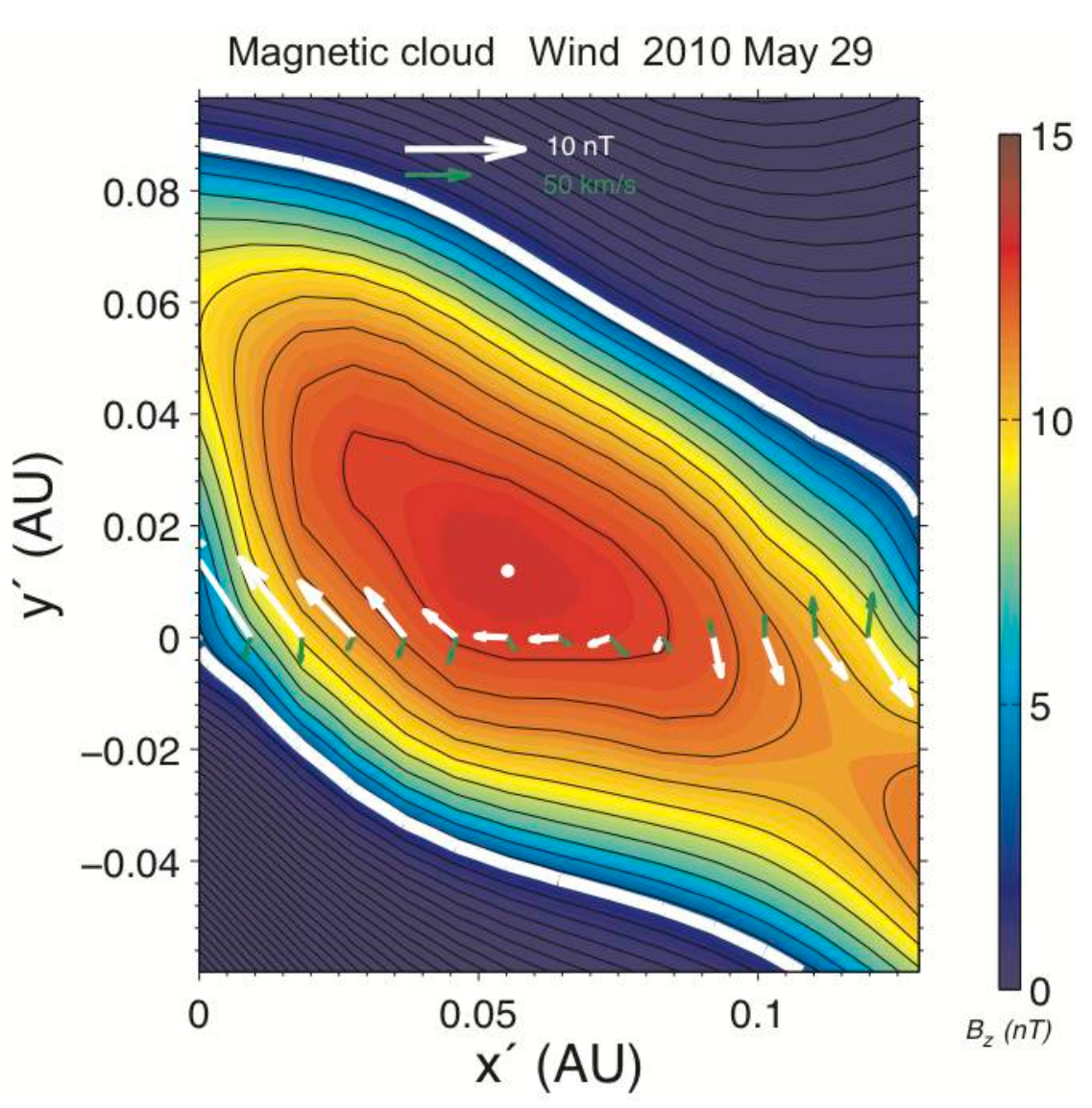}}
\end{minipage}
\vspace{0.08cm}
\caption{{\it In situ} data as observed by the {\it Wind} spacecraft ({\it left}) from May 27 20 UT to May 30 08 UT and zoom-in on the ejecta and the last part of the sheath after minimum-variance analysis ({\it top right}). The bottom right panel shows the Grad-Shafranov reconstructed cross-section map (see text for details). The black contours show magnetic field lines in a plane perpendicular to the flux rope axis, which is indicated by a white dot; and the axial field strength is color coded. Along the {\it Wind} trajectory, remaining velocities in the deHoffmann-Teller frame are indicated by green arrows, and observed magnetic field components in the plane of the cross-section are shown with white arrows.}
\end{figure*}

Overall, images and J-maps from HI2 strongly suggests that the first track in HI2-A corresponds to CME2, which is at a larger heliocentric distance and closer to STEREO-A than CME1. We can hypothesize that the ``split'' of the first track in HI2-A, which merges with the second track, occurs at the time when the elongation angle of CME2 becomes equal and then larger than that of CME1. 


\section{Arrival Time at 1~AU and {\it In Situ} Measurements} \label{insitu}

\subsection{Overview of the {\it In Situ} Measurements}

Magnetic field and plasma data from the Wind spacecraft are shown in the left column of Figure~8 for the period 20 UT, May 27 to 8 UT, May 30, 2010.
{\it Wind} was orbiting the L1 Lagrangian point and at the start and end of the 
period it was at (262.2, -13.0, 25.3) $R_E$ and (261.9, -18.2, 25.2) $R_E$, respectively (GSE coordinates). The plasma data are from the Three-Dimensional Plasma Analyzer \citep[]{Lin:1995}, and the magnetic field data are from the Magnetic Field Investigation instrument \citep[]{Lepping:1995}. Both have a  temporal resolution of three seconds. The panels show, from top to bottom, the proton plasma
density (in red, the alpha particle-to-proton number density ratio in percent),
the proton temperature (in red, the expected proton temperature for
normal solar wind expansion after \citet{Lopez:1987}), the bulk flow
speed and the east-west component ($V_y$) of flow vector, the 
pressures (total, $P_t$, magnetic, $P_m$, and the sum of the 
proton and electron thermal pressure, $P_p + P_e$), the
strength and GSE components of the magnetic field, $\bf{B}$, colored as shown, and the proton plasma $\beta$. 

The data reveal a clear, single interplanetary coronal mass ejection
passing {\it Wind} at the times shown between the second and third vertical guidelines.
It is characterized by a strong field, smooth rotations of $\bf{B}$ components,
and a low proton $\beta$. The alpha particle-to-proton density ratio is
generally also higher than average.  The proton temperature is comparable
and even above the expected proton temperature, so the 
configuration is not a magnetic cloud \citep[]{Burlaga:1981}. The rotation 
of the field indicates, however, that it is a magnetic flux rope (see further
below).

A shock (marked as S) precedes the ICME  by $\sim$18 hrs. We examined this shock 
by the velocity coplanarity method \citep[]{Abraham:1972} and 
obtained a shock speed of 392 km~s$^{-1}$. This is comparable with the 
speed at the leading edge of the ICME, consistent with it being driven
by the transient. The angle between the upstream magnetic field and
the shock normal is 14.2$^{\circ}$, hence it is a quasi-parallel
shock. 
While it may be surprising that such a slow magnetic cloud drives a shock, it should be noted that the Alfv{\'e}nic and sonic speeds upstream of the shock are both below 40~km~s$^{-1}$ so that the CME is faster than the fast magnetosonic speed in the solar wind frame. The sheath preceding the CME is found to be about 0.16~AU long. 


\begin{table*}[tb*]
\centering
\begin{tabular}{cccccc}
\hline
\hline
Front &  Method & Direction & Arrival Time & Speed at 1~AU\\
\hline
{\bf HI2-A Track 1} & F$\Phi$ & 5$^\circ$ & 05:00UT on 05/27 & 630~km~s$^{-1}$\\
{\bf HI2-A Track 1} & F$\Phi$ & 12$^\circ$ & 13:00UT on 05/27 & 530~km~s$^{-1}$\\
{\bf HI2-A Track 1} & HM & 24$^\circ$ & 09:00UT on 05/28 & 370~km~s$^{-1}$\\
\hline
{\bf HI2-A Track 2} & F$\Phi$ & 5$^\circ$ & 03:00UT on 05/28 & 490~km~s$^{-1}$\\
{\bf HI2-A Track 2} & HM & 5$^\circ$ & 11:00UT on 05/28 & 390~km~s$^{-1}$\\
\hline
{\bf HI2-B Track} & F$\Phi$  & 5$^\circ$ & 04:00UT on 05/28 & 390~km~s$^{-1}$\\
{\bf HI2-B Track} & HM & 24$^\circ$ & 09:00UT on 05/28 & 360~km~s$^{-1}$\\
\hline
{\bf HI1-A LE1} & F$\Phi$ &  & 05:30UT on 05/28 & 370~km~s$^{-1}$\\
{\bf HI1-A BC1} & F$\Phi$ &  & 19:00UT on 05/28 & 340~km~s$^{-1}$\\
{\bf HI1-A LE2} & HM & 28$^\circ$ & 23:00UT on 05/28 & 345~km~s$^{-1}$\\
\hline
{\bf Shock} & {\it Wind} & & 02:00UT on 05/28 & 390~km~s$^{-1}$\\ 
{\bf End of Sheath - Start Ejecta} & {\it Wind}& & 20:00UT on 05/28 & 390~km~s$^{-1}$\\
{\bf Compression - End Ejecta} & {\it Wind} & & 20:00UT on 05/29 & 360~km~s$^{-1}$\\
\hline
\hline
\end{tabular}
\caption{Predicted and measured arrival times and speed of the bright tracks observed by SECCHI on May 23-27, 2010 and {\it in situ} by {\it Wind}. LE refers to leading edge and BC to the beginning of the cavity. The predictions using HI1-A data assume a constant speed in HI-2 (see section 5.2 for details).}
\end{table*}

We examined the ICME magnetic field data by a minimum variance technique
\citep[e.g.][]{Sonnerup:1967} to confirm the visual inspection. With a ratio of intermediate-to-minimum eigenvalues of 8.1, the routine returns a 
reliable result. 
As shown in the top right panel of Figure~8, the flux rope axis points predominantly in the GSE $z$-direction, with a small tilt in the $y$-direction (as is clear from the direction of the k vector). This is consistent with the large tilt out-of-the-ecliptic found with the GCS fitting.
For an ICME with such an orientation, it is expected that the cross-section in the ecliptic plane is relatively small and the cloud can be detected at Earth only if it propagates very close to the Sun-Earth line. The presence of a non-zero component of $B_i$ inside the cloud indicates a non-zero impact parameter. Considering the cloud is left-handed, this means that the center of the ejecta passed slightly west of the Sun-Earth line (meaning it passed on the $-y$ side of Earth in GSE coordinates). The orientation of the shock normal (0.90, -0.40, -0.15) in GSE coordinates is also consistent with this picture.

Grad-Shafranov reconstruction \citep[]{Hu:2001,Moestl:2009} was also performed, and it also yields a single, left-handed ejecta, with an axis making an angle of 10$^\circ$ with respect to the $-z$ direction with a small impact parameter of 0.1 relative to the radial size of the flux rope. The cross-section map is shown in the bottom right panel of Figure~8. The view is approximately from ecliptic south with the Sun on the right-hand side of the picture. The Grad-Shafranov reconstruction technique has been shown to work best for magnetic clouds and magnetic cloud-like ejecta (as compared to less regular ejecta, see \citet{AlHaddad:2012} for details) and this shows once again that the {\it in situ} measurements indicate a relatively typical flux rope-type ejecta, except for the higher plasma temperature. The axial field strength is found to be 13.5~nT, which is typical for a magnetic cloud or magnetic cloud-like ejecta at solar minimum. 

Returning to the left column of Figure~8, one may notice that before the 
ICME is crossed, the sheath density decreases steadily while  the magnetic field strength
increases. These are features similar to those of a plasma depletion layer.
A flow enhancement consistent with this expectation
is shown in Figure~8, left column third panel, at the location of the second vertical 
guideline. In this interpretation, it is a draping effect. Draping around ICMEs observed at Earth has been described previously \citep[e.g., see][]{Liu:2006}.


\subsection{Prediction from HI Observations}

We use the measurements in HI2 combined with various estimates of the direction of the CMEs to calculate the arrival times at Earth of the bright fronts. The results are summarized in Table~1. 
Using the F$\Phi$ approximation and a direction of 5$^\circ$ to 12$^\circ$ for the first track in HI2-A, the predicted arrival time is 12 to 20 hours earlier than any structure detected by {\it Wind}. This confirms that the first track in HI2-A is not associated with CME1. In order to obtain a consistent arrival time and speed at 1~AU for HI2-A and HI2-B data, it is necessary to consider that the first track in HI2-A and the track observed in HI2-B correspond to CME2 with a direction $\sim 24^\circ$. This substantiates the discussion in section 4.5, where we found that the first track observed in HI2-A corresponds to CME2. It should be noted that a CME passing 24$^\circ$ off the Sun-Earth line is not expected to result in the type of clear flux rope measurements as seen in Figure~8, especially for a CME with such a large tilt with respect to the ecliptic.

As explained in section 4.5 and further confirmed above, the second track in HI2-A corresponds to CME1. Assuming a direction of 5$^\circ$ for CME1 (due to the deflection during the interaction, see section 6 for more details), the arrival time at {\it Wind} can be well reproduced using the F$\Phi$ approximation, but  the predicted arrival speed is too high by more than 200~km~s$^{-1}$. The opposite is true with the HM approximation (arrival time too late by 7 hours but speed well reproduced). This might be because CME1 has a narrow, but non-negligible width, and, also because at large elongation angles, neither method works well. 

We also use the observations in HI1 to predict the arrival times of CME1 and CME2 without taking into account any additional acceleration or deceleration during the remaining of the CME-CME interaction. We simply assume that, past 0.35~AU, the CMEs propagate with constant speeds equal to the final speed determined from HI1 data. Results are also summarized in Table~1. Assuming a constant speed of 370~km~s$^{-1}$,  the leading edge of CME1 is predicted to arrive at 1~AU within four hours of the measured shock arrival time. The predicted arrival time for the magnetic ejecta is within two hours of the start of the magnetic cloud observed by {\it Wind}.

Using the expansion rate observed before the CME-CME collision, CME1 is expected to arrive at 1~AU with a radius of about 0.11~AU, which is similar to the measured radius of 0.1~AU. The same procedure yields an expected thickness of about 0.1~AU for the sheath in front of CME1, a slight underestimation of the measured sheath thickness.

\section{Discussion: Scenario of the CME-CME interaction}\label{discuss}

Here, we attempt to explain simultaneously the {\it in situ} measurements showing a single ICME with a flux-rope type structure propagating close but to the west of the Sun-Earth line and the remote-sensing observations showing the interaction of two CMEs.  To do so, it is necessary to consider that, when the two CMEs interact, they are deflected away from each other, which ultimately results in the two CMEs separating from each other. 

As determined in sections 3.2 and 4.3, CME1 has an initial direction of about 12$^\circ$ and a half-width of about 10$^\circ$. CME2 has an initial direction of about 28$^\circ$ (see section 3.2). The collision between the two CMEs involves a relatively small cross-section (as sketched in the bottom panel of Figure~5). As the two CMEs collide, there is an eastward deflection of CME1 and westward deflection of CME2. CME1 is originally propagating along W12 and is observed at 1~AU with a very small impact parameter, consistent with a small western direction. Therefore, we can infer that the deflection of CME1 is of the order of 8-10$^\circ$.
The only numerical study of CME deflection during CME-CME interaction was performed by \citet{Xiong:2009}, focusing only on the latitudinal deflection. The authors found a maximum deflection when the two CMEs propagate 15$^\circ$ away from each other, as is the case here (see their Figure 12c). However, the deflection was found to be only 3-6$^\circ$ for their CME1.

Initially, CME1 is compressed by the CME-CME interaction; because of the deflections, the two CMEs separate and CME1 is ``free'' to expand again. It has been previously proposed \citep[see for example Figure 6 of ][]{Gulisano:2010} that a compressed magnetic cloud may over-expand to reach a size similar to what would be its size in the absence of the compression event. Compression may be due to a fast wind solar wind stream overtaking the magnetic cloud or an instance of CME-CME interaction. We believe that such an over-expansion is possible only if there is no other magnetic obstacle (another CME) at the back of CME1 to hinder it. Therefore, CME1 and CME2 must separate after the collision so that the magnetic ejecta associated with CME1 over-expands to its expected size (that given by the relation found in section 4.3). The over-expansion results in CME1 having a typical radius at 1~AU and no {\it in situ} signatures of its compression, except, possibly, the higher proton temperature being present. This is also the reason why the measured width of CME1 at 1~AU is consistent with the prediction from HI1 data only.

Due to its eastward deflection, {\it Wind} observes CME1, including its sheath and associated magnetic flux rope structure. CME2 is not observed {\it in situ}, having been deflected further to the west. After the data gaps in SECCHI, CME1 is observed in HI2-A as the second track and in HI2-B as the first (and only) track. CME2 is observed in HI2-A as the first track, which shows a geometrical acceleration consistent with a direction closer to STEREO-A than CME1. During the first 12 hours of propagation in HI2-A, there is only one bright feature observed in background-subtracted images. During this time, the signals from the two CMEs are  superimposed and the track associated with CME2 ``passes'' that associated with CME1. Because of the running-difference procedure, the tracks appear continuous in the classical J-map format, except for the ``split'' in the first track. 

\section{Conclusions} \label{conclusion}

In this paper, we have analyzed in detail the remote-sensing observations of two successive CMEs in the inner heliosphere and the associated {\it in situ} measurements. The two CMEs were associated with eruptions from the same filament channel and they are clearly observed to interact in SECCHI/HI1-A images. The interaction is associated with a simultaneous decrease in speed of the second CME, and a compression in the radial direction of the first CME. We have shown how, due to the two CMEs having different initial directions (by about 10-15$^\circ$), CME1 is deflected towards the Sun-Earth line (eastward) and CME2 away from Earth. At 1~AU, only CME1 is observed as an ICME with a radius of 0.11~AU and preceded by a sheath of thickness 0.15~AU. The ICME has the geometry of a flux rope with an axis along the $-z$-direction and all the characteristics of a magnetic cloud except an elevated plasma temperature. While the sheath has interesting features, including some evidence of draping around the ICME, there is no hint of a past CME-CME interaction in the {\it in situ} measurements, except perhaps for its high density and long (but not unusual) duration. 

In our study of remote-sensing observations, we have made, or confirmed, a number of other findings.  1) We have determined that the radial expansion of the magnetic ejecta with distance in the inner heliosphere can be approximated by a power law of exponent 0.82 \citep[similar to the studies by][]{Bothmer:1994,Liu:2005} and there is a weaker radial dependence for the sheath thickness.  2) We have found that the radius of the previously perturbed ICME at 1~AU is consistent with the expansion rate calculated in the inner heliosphere between 0.1 and 0.35~AU (before the collision), confirming previous studies of compression of magnetic clouds by solar wind streams and other magnetic clouds \citep[]{Gulisano:2010}. 3) We have shown that each part of a CME observed in HI images can be assumed to propagate in the same direction, including CME tracks which are created through the running-difference procedure (for example, the ``black'' edge). 4) We have derived the temporal offset between the actual leading edge of a CME as seen in white-light images and the ``black'' edge as seen in running-difference images. This black edge is what is typically tracked in HI2 and used to estimate the CME arrival times from fitting techniques such as those from \citet{Rouillard:2008} and \citet{Lugaz:2010c}. For the 2010 May 23 CME, we found that the temporal offset increases with time and is approximately 2-3 hours in HI1 and 4-6 hours in HI2.

This study is also meant to demonstrate the difficulties of relying solely on remote-sensing observations to predict complex ICME events at Earth. While heliospheric imagers have improved the prediction of arrival times and arrival speed of CMEs \citep[e.g., see][]{Davis:2011, Liu:2010b, Schreiner:2012}, we are still struggling to understand complex events, such as the August 2010 series of CMEs or this series of two CMEs in May 2010. Here, 
for example, the tracks associated with the two CMEs do not appear to merge or cross with each other in the J-map made with STEREO-A data, whereas, in fact, the second, faster CME collided with the first, slower CME and may be ahead of it (but along a different direction) at 1~AU. In addition, there is only one bright track in HI2-B 

During the interaction, the speed of CME2 is found to decrease from close to 600~km~s$^{-1}$ to about 380~km~s$^{-1}$. Comparing its speed in the HI1 FOV with the speed measured at 1~AU, it is clear that CME1 is only slightly accelerated during the collision. It would appear that the changes in the speeds of the two CMEs are consistent with a perfectly inelastic collision, but a more dedicated study is required to analyze in details the type of interaction. Such a study should take into consideration the expansion of the CMEs, their deflection, the internal magnetic field in the two CMEs, the reconnection between the CMEs, and the possible presence of a shock ahead of CME2. Similarly to what was determined in the numerical simulation of \citet{Lugaz:2005b}, we find that, following the collision, the front of CME2 propagates with a speed comparable with, but slightly higher than the speed of the front of CME1. This results in a contraction of CME1 after the collision, as its back moves faster than its front. We speculate that the speed of CME2 after the collision might be constrained by the speed of the leading edge of CME1 because the magnetic tension and pressure inside CME1 limit the rate of contraction of CME1. In this interpretation, CME1 contraction speed constrains the speed of the front of CME2 to be close to, but slightly higher than the speed of the front of CME1.

In the future, it will be important to perform more dedicated numerical studies of the interaction of two (or more) CMEs with different orientations, sizes and longitudinal (and latitudinal) separations in three dimensions, with high resolution and including synthetic HI images, a combination of the detailed simulation of \citet{Lugaz:2005b} with the cases studied in \citet{Xiong:2009}.

\begin{acknowledgments}
The authors would like to thank the reviewer for his/her useful comments which have helped make our manuscript clearer.
The research for this manuscript was supported by the following grants: NASA NAS5-00132, NNX10AQ29G, and NNX12AB28G and NSF AGS-1239699, and AGS-1239704. 
I.~R. would like to acknowledge the support from CAS grant 2011T2J01 at the YNAO and NSF grant AGS-0639335 (CAREER) at the IfA.
M.~T. acknowledges the Austrian Science Fund (FWF): V195-N16. C.~M. acknowledges support from a Marie Curie International Outgoing Fellowship within EU FP7 and from the EU FP7 program under grant agreement n$^{\circ}$263252 [COMESEP].
SoHO and {\it STEREO} are projects of international cooperation between ESA and NASA. 
The SECCHI data are produced by an international consortium of 
  NRL, LMSAL, and NASA GSFC (USA), RAL, and U.
  Birmingham (UK), MPS (Germany), CSL (Belgium), IOTA, and IAS (France).  
\end{acknowledgments}

\bibliographystyle{apj}


\end{document}